\documentclass[aps,prb,twocolumn,superscriptaddress, showpacs]{revtex4-2}

\usepackage{graphicx}% Include figure files
\usepackage{dcolumn}% Align table columns on decimal point
\usepackage{bm}% bold math
\usepackage{hyperref}% add hypertext capabilities
\usepackage{bm, amsmath}
\usepackage{txfonts}
\usepackage{xcolor}

\begin{document}

%\title{Stability and collective dynamics of skyrmionium-based meta-matter in quasi-two-dimensional chiral magnets}
\title{Attractive Hopfions and Bimerons in Thin Films of Chiral Magnets: Cluster Formation and Lattice Instability in the Conical Phase}

\author{Andrey O. Leonov}
\thanks{Corresponding author: leonov@hiroshima-u.ac.jp}
\affiliation{Department of Chemistry, Faculty of Science, Hiroshima University Kagamiyama, Higashi Hiroshima, Hiroshima 739-8526, Japan}
\affiliation{International Institute for Sustainability with Knotted Chiral Meta Matter (WPI-SKCM$^2$), Hiroshima University, 1-3-1 Kagamiyama, Higashi-Hiroshima, Hiroshima 739-8531, Japan} 

\author{Takayuki Shigenaga}
%\thanks{leonov@hiroshima-u.ac.jp}
\affiliation{Department of Chemistry, Faculty of Science, Hiroshima University Kagamiyama, Higashi Hiroshima, Hiroshima 739-8526, Japan}
\affiliation{International Institute for Sustainability with Knotted Chiral Meta Matter (WPI-SKCM$^2$), Hiroshima University, 1-3-1 Kagamiyama, Higashi-Hiroshima, Hiroshima 739-8531, Japan}

\date{\today}

\begin{abstract}
We investigate the energetics, interactions, and ordering tendencies of bimerons (cholesteric fingers of the second type, CF-2) and hopfions in thin films of chiral magnets and chiral liquid crystals hosting a conical background state. Although isolated bimerons possess positive eigen-energy with respect to the conical phase, they develop an attractive interaction mediated by the restructuring and partial overlap of their positive-energy shells—intermediate regions formed relative to the conical state. This attraction promotes the formation of bound pairs and extended bimeron chains, even in parameter regimes where a periodic bimeron lattice is no longer thermodynamically stable.

Extending the analysis to three dimensions, we show that circularization of bimerons into hopfions renders their energy finite and gives rise to a well-defined metastability window closely linked to the stability range of cholesteric fingers. Isolated hopfions likewise exhibit an attractive interaction within the conical phase, leading to the formation of hexagonally ordered clusters. The attraction originates from the competition between favorable and unfavorable twist regions and from the energetic cost of the shell structures imposed by the conical background.

Despite the presence of attractive pair potentials and cluster formation, we demonstrate that hexagonal hopfion lattices do not exhibit an equilibrium lattice period. Instead, the system evolves toward states in which the conical spiral or the CF--1 phase (cholesteric fingers of the first type) progressively invade the inter-soliton regions, thereby preventing crystallization. Our results reveal a regime of attraction without stable long-range order and clarify the interplay between topology, confinement, and conical-phase frustration in chiral-magnet and liquid-crystal thin films.
\end{abstract}

\maketitle

\section{Introduction}

Topological solitons are spatially localized, particle-like configurations of continuous fields that arise across diverse physical systems, from nuclear matter to condensed media. Their stability is protected by topological invariants, which impose energetic barriers separating these nontrivial textures from the surrounding  (“vacuum”) states~\cite{manton_sutcliffe,shnir,Volovik,solitons}.
Although often treated as isolated excitations, solitons generally interact, exhibiting either attraction or repulsion and forming ordered crystalline and/or clustered states.
Such self-organized assemblies can extend over mesoscopic scales much larger than a single core. We refer to these collective states as \emph{solitonic meta-matter}, where the effective building blocks are topologically protected textures acting as emergent particles~\cite{leonov2026metamatter}.

Among the many realizations of topological solitons and their associated meta-matter,
hopfions have emerged as paradigmatic examples in condensed-matter systems, including chiral magnets (ChM) and chiral liquid crystals (CLC). 

From a topological viewpoint, hopfionic textures are maps
\(
\mathbf{m} : \mathbb{R}^3 \cup \{\infty\} \simeq S^3 \rightarrow S^2
\),
classified by the third homotopy group~\cite{Faddeev,Bott},
\(
\pi_3(S^2) \cong \mathbb{Z}
\).
This homotopy class is labeled by an integer-valued Hopf invariant, which admits a geometric interpretation as the linking number of preimages of two generic points on the order-parameter sphere. Equivalently, it measures the total linking of the field lines associated with the configuration,
\begin{equation}
    Q_H = \frac{1}{(4\pi)^2} \int_{\mathbb{R}^3} \mathbf{A} \cdot \mathbf{B}\, \mathrm{d}^3 x,
\end{equation}
where the emergent field
\(
\mathbf{B}
\)
is defined as
\begin{equation}
\mathbf{B}_i
=
\frac{1}{2}
\epsilon_{ijk}\,
\mathbf{m}\cdot
\left(
\partial_j \mathbf{m}
\times
\partial_k \mathbf{m}
\right),
\end{equation}
and satisfies
\(
\nabla \cdot \mathbf{B} = 0.
\)
Therefore there exists a vector potential \(\mathbf{A}\) such that $\nabla \times \mathbf{A} = \mathbf{B}$.
Here \(\mathbf{m}\) is a smooth unit vector field, interpreted physically as the magnetization in ChM or the director field in CLC.

On the other hand, hopfions can be viewed as stable, spatially localized solutions of suitable phenomenological continuum theories. In chiral magnets, the theoretical framework describing skyrmions and other nontrivial magnetization textures was originally developed by Dzyaloshinskii~\cite{Dz64} [see Eq.~\ref{functional}]. In its simplest isotropic form, the corresponding energy functional $W(\mathbf{m})$ includes the exchange interaction and the Dzyaloshinskii--Moriya interaction (DMI), supplemented by a Zeeman coupling to an external magnetic field $\mathbf{H}$. The DMI~\cite{Dz64,moriya} arises from spin--orbit coupling and represents an antisymmetric exchange between neighboring spins that energetically favors twisted magnetic configurations, thereby providing their internal stability. In CLC, analogous DMI-like terms arise from the acentric shape of the constituent molecules and, within the Frank–Oseen phenomenological framework \cite{Oswald,kleman2003}, similarly promote the formation of solitonic textures.

Nowadays, three-dimensional magnetic hopfions attract considerable interest due to their potential applications, e.g., in spintronic devices, driven by their emergent electromagnetic response and nontrivial dynamical behavior under external stimuli. In CLC, hopfions are promising for reconfigurable photonic and optoelectronic applications, as they produce controllable three-dimensional refractive-index modulations and can be created, manipulated, and erased by optical or electric fields \cite{ackerman2017}.

Hopfions have recently been observed in magnetic~\cite{Kent}, ferroelectric~\cite{luk2020hopfions}, and liquid-crystalline systems~\cite{ackerman2017,ackerman2017static}, and have also been investigated in Bose–Einstein condensates~\cite{bidasyuk2015stable}.
In most realizations, hopfion stability requires specially engineered conditions. In frustrated magnets~\cite{sutcliffe2018hopfions,sallermann2023stability} with competing exchange interactions, both rotational senses of the magnetization are energetically equivalent, which favors the formation of hopfions. In magnetic nanostructures~\cite{liu2018lake} and multilayers~\cite{Kent}, perpendicular magnetic anisotropy can be tuned to introduce an energy barrier that prevents hopfion collapse. In CLC, the disparity of elastic constants together with surface anchoring promotes hopfion stabilization.
In the context of hopfions, confinement is a crucial prerequisite for their existence. Indeed, in bulk chiral magnets, hopfions tend to elongate along the direction of the applied magnetic field and ultimately relax into spiral states~\cite{leonov2023swirling,metlov2025elliptical}. In contrast, thin-film geometries impose finite-size constraints and boundary conditions that suppress such elongation and profoundly reshape the topological phase landscape. In general, hopfions are stabilized in confined systems with a film thickness of the order of one spiral pitch.

Hopfionic quasiparticles may presumably self-organize into extended meta-matter in the form of a hopfion lattice (HL) or a hopfion crystal. The formation of such collective states is of particular interest, as it reflects the interplay between topology, symmetry, and interactions in complex systems and can generate emergent phenomena and collective modes that are absent at the level of individual solitons.

By analogy with 2D skyrmions, a spontaneous condensation of hopfions is expected when the eigen-energy of an isolated hopfion drops below that of the homogeneous  background phase, leading to a periodic array with an equilibrium inter-hopfion spacing~\cite{Bogdanov94}. An apparently stable hopfion lattice was reported in Ref.~\cite{tai2018} within the same phenomenological model~(\ref{functional}) considered in the present work. In Refs.~ \cite{hopfions,leonov2026precursor}, however, it was argued—based on general considerations and numerical simulations—that the HL is intrinsically unstable.
This instability was attributed to the internal structure of an isolated hopfion, which can be regarded as a closed, circular realization of a bimeron texture. Once the energy of a bimeron becomes negative with respect to the homogeneous state, a modulated bimeron phase can densely fill space without leaving voids. In contrast, a hopfion lattice necessarily contains extended regions of homogeneous ``vacuum,'' both within the hopfion cores and in the interstitial regions between neighboring objects. This geometric inefficiency favors a deformation of the hopfion meta-matter, driving the collapse of the vacuum cores and their subsequent elongation into bimeron-like structures \cite{hopfions}. %Such behavior is also consistent with the form of the inter-bimeron interaction potential, which determines an equilibrium separation between planar bimerons. 
A similar meta-matter instability was recently reported also for skyrmioniums in quasi-two-dimensional chiral magnets~\cite{leonov2026metamatter,Nakamura}, where both hexagonal and square skyrmionium lattices were found to elongate and eventually transform into spiral states. This parallel is not coincidental: skyrmioniums constitute the natural two-dimensional cross-sections of hopfions, and their behavior therefore reflects key features of the underlying three-dimensional topology.

{\color{black}An additional conclusion of Refs.~\cite{hopfions,leonov2026precursor} was that the metastability region of isolated hopfions within the homogeneous state is closely related to the stability region of the corresponding finger phases (CF--2). In this sense, hopfions were interpreted as precursor states of cholesteric fingers. In Refs.~\cite{hopfions,leonov2026precursor}, bulk uniaxial anisotropy served as the control parameter stabilizing the homogeneous background state.
%In Ref. \cite{hopfions}, isolated hopfions were investigated in the presence of bulk uniaxial anisotropy, which stabilized the homogeneous background state. 

In the present manuscript, by contrast, we investigate a system without bulk uniaxial anisotropy and employ the applied magnetic field as the principal control parameter. This setting corresponds to the minimal Dzyaloshinskii model supplemented by interfacial perpendicular magnetic anisotropy (PMA). 
Under these conditions, the finger phases emerge within the stability region of the conical phase rather than within the homogeneous state. Consequently, our goal is to determine whether the conclusions drawn in Refs.~\cite{hopfions,leonov2026precursor}—namely, the metastability of isolated hopfions and the instability of the hopfion lattice (HL)—remain valid in the presence of the conical background, or whether the conical phase qualitatively modifies hopfion energetics, interactions, and stability.}

%the relevant problem becomes the stability of isolated hopfions embedded in a conical background and the stability of hopfion lattices. 

%In the present work, we investigate the stability of hopfions within the minimal Dzyaloshinskii model supplemented by interfacial perpendicular magnetic anisotropy (PMA). In this framework, hopfions are confined to thin films with a thickness equal to one spiral pitch. As the background state, we consider the conical phase rather than the homogeneous state. 
%
%Our goal is to determine whether the conclusions drawn in Ref.~\cite{hopfions}—namely, the metastability of isolated hopfions and the instability of the hopfion lattice—remain valid in this setting, or whether the conical background fundamentally modifies hopfion energetics and stability.

As a first result, we demonstrate that isolated hopfions also become stable when embedded in a surrounding conical state. In this configuration, the hopfion forms an energetically costly transitional region with respect to the conical background (hereafter referred to as a ``shell,'' in analogy with the intermediate region between skyrmions and the conical phase introduced in Ref.~\cite{leonov2022skyrmion}). 
Consequently, an eigen-energy minimum with respect to the hopfion radius emerges naturally. Increasing the hopfion radius beyond its equilibrium value allows the conical phase to penetrate into the interior, leading to the formation of an additional internal shell and thereby increasing the total energy. Conversely, decreasing the hopfion radius reduces the negative energy contribution associated with the nearly homogeneous state occupying the hopfion core. The competition between these effects stabilizes the hopfion at a finite equilibrium size.
These findings clarify the role of the conical phase: rather than suppressing hopfions, it provides a structured background that qualitatively modifies their energetics through the formation of shell regions and thereby governs their equilibrium properties. %Consequently, while the overall picture of isolated-hopfion metastability persists, the presence of the conical state introduces additional energetic contributions that govern their size, interaction, and collective behavior.

Second, the formation of the shell also underlies the attractive inter-hopfion interaction potential and promotes clustering, giving rise to extended hopfion conglomerates.
When two hopfions approach each other, their respective shell regions overlap and partially merge, thereby reducing the total area of energetically costly transitional regions. As a result, the system lowers its total energy by minimizing the combined shell volume, which manifests as an effective short-range attraction between hopfions.
In extended assemblies, this attraction favors the formation of compact clusters while preserving, to a large extent, the underlying triangular ordering that corresponds to the densest packing in two dimensions.  
Thus, the shell is not merely a passive boundary feature but constitutes the central energetic ingredient that governs both the stability of individual hopfions and the emergence of collective states in the conical background.

In contrast, an extended hopfion lattice—viewed as an infinite cluster without boundaries—is unstable and transforms into other, more energetically favorable modulated phases. In particular, the absence of boundary-induced shell reduction prevents the lattice from gaining the energetic advantage that stabilizes finite clusters. As a consequence, the accumulated shell energy in the bulk of the lattice renders such periodic arrangements unfavorable, and the system relaxes toward alternative chiral modulated states, such as the conical or CF-1 phases, which provide a lower average energy density.
We argue that this result resolves the apparent contradiction between experimentally reported ``hopfion lattices'' in CLC~\cite{ackerman2017static}—which may, in fact, correspond to finite hopfion clusters stabilized and confined by the surrounding conical background—and the theoretical predictions of the present work and Ref.~\cite{hopfions}, which consistently indicate the instability of an ideal, infinite hopfion lattice in the absence of boundary-induced stabilization mechanisms.

Furthermore, as hopfions are intimately connected to the conical and cholesteric finger phases, we provide a magnetic perspective on these configurations known in CLC and analyze their internal structure and properties in detail.

%In this work, we perform numerical simulations to predict stable static hopfions in noncentrosymmetric magnetic nanostructures with interfacial perpendicular magnetic anisotropy (PMA). We show that, in addition to Dzyaloshinskii--Moriya interactions (DMI)~\cite{10}, confinement and interfacial PMA play a crucial role in stabilizing hopfions. Our study focuses on fully nonsingular field configurations.

\begin{figure*}
  \centering
  \includegraphics[width=0.9\linewidth]{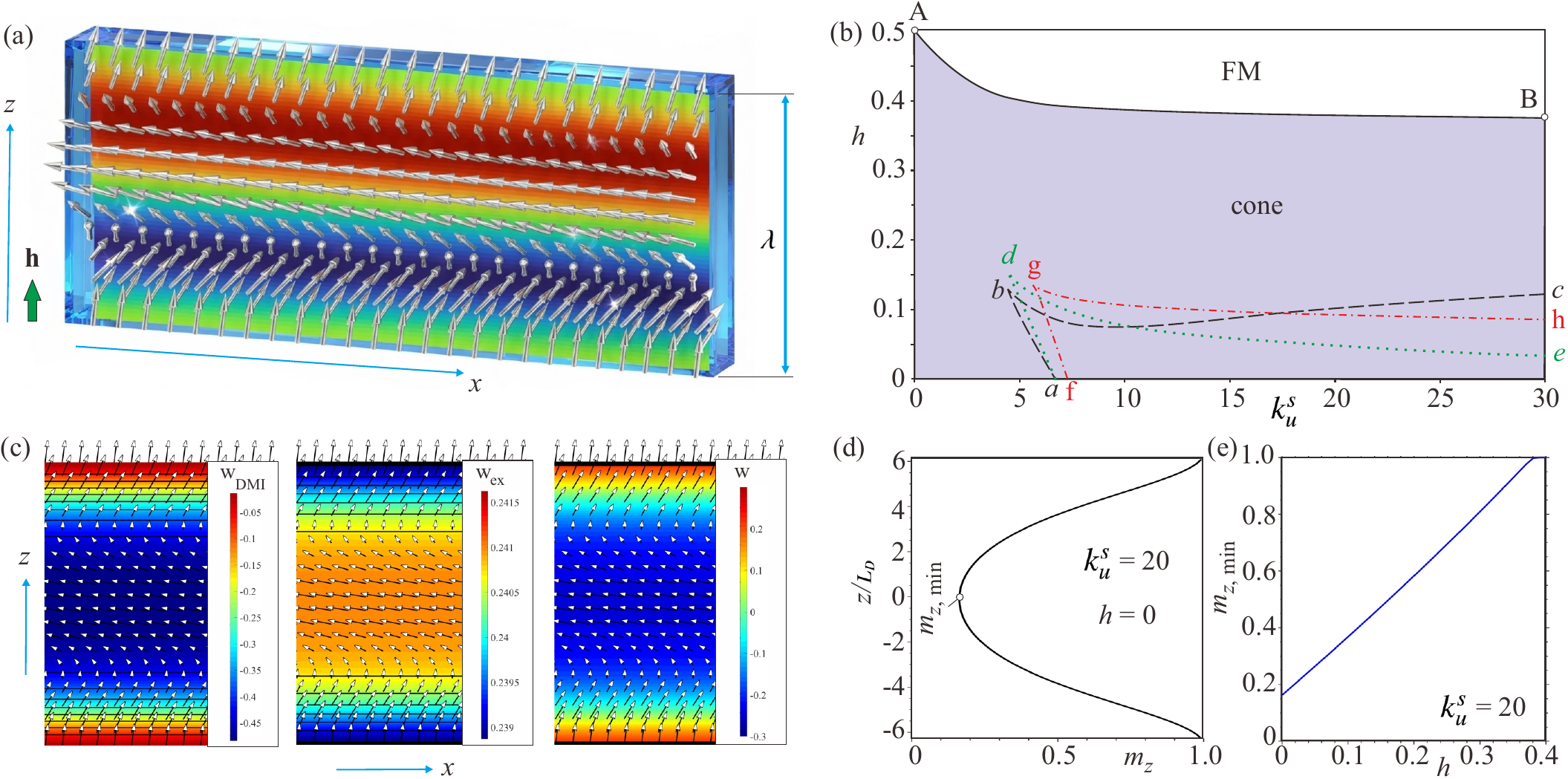}
 \caption{\label{fig01}
(Color online) Confined conical state in a thin film of thickness equal to one spiral pitch, $\lambda = 4\pi$. 
(a) Schematic illustration of the magnetization distribution within the conical spiral shown by gray arrows, together with a color map of the $m_y$ component using the standard color scale from $-1$ (blue) to $+1$ (red). 
(b) Phase diagram in the $(h, k_u^s)$ plane. The line $A$--$B$ denotes the second-order transition from the conical to the homogeneous state; point $A$ corresponds to $\theta_0 = 0$ and the critical field $h = 1/2$, in agreement with the analytical solution~(\ref{anal}). {\color{black} The curvilinear boundary $a$--$b$--$c$ (black dashed line) separates the parameter region where the CF--1 phase becomes energetically favorable with respect to the conical phase (see Sect.~\ref{sect:CF1stability}). The metastability domain of isolated bimerons is indicated by the boundary $a$--$d$--$e$ (green dotted line) (see Sect.~\ref{sect:CF2stability}). The region enclosed by the boundary $f$--$g$--$h$ (red dash-dotted line) corresponds to metastable hopfions embedded in the conical background (see Sect.~\ref{sect:Hopfstability}).}
(c) Color maps of the energy density distribution across the film thickness for $h = 0$ and $k_u^s = 20$: DMI energy density (left), exchange energy density (center), and total energy density (right). 
(d) Magnetization profile $m_z(z)$ across the layer, showing nearly saturated spins at the surfaces and a minimum in the center of the film. 
(e) Field dependence of the minimal magnetization component $m_{z,\min}$; the transition to the homogeneous state occurs when $m_{z,\min} = 1$. The corresponding maximal cone angle $\theta_{\max} = \arccos(m_{z,\min})$ is used in Eq.~(\ref{thetamax}).
}
\end{figure*}

%%%%%%%%%%%%%%%%%%%%%%%%%%%%%%%%%%%%%%%%%%%%%%%%%%%%%%%%%%%%%%%%%%%%%%%%%%%
\section{Phenomenological model\label{sect:model}}

Within the Dzyaloshinskii phenomenology~\cite{Dz64}, the total energy of a noncentrosymmetric ferromagnet is given by~\cite{leonov2014,leonov2021}
\begin{align}
W(\mathbf{m}) 
&= \int_{\Omega} \Big[
A (\nabla \mathbf{m})^2
+ D \, \mathbf{m} \cdot (\nabla \times \mathbf{m})
-\mu_0 M_s \mathbf{H} \cdot \mathbf{m}
\Big] \, d^3 \mathbf{r}\nonumber\\
& \quad - \int_{\partial \Omega} K_s (\mathbf{m} \cdot \mathbf{z})^2 \, d^2 \mathbf{r},
\label{functional}
\end{align}
where \(A\) is the exchange stiffness, \(D\) is the DMI constant. The unit vector \(\mathbf{m}(\mathbf{r})\) describes the local magnetization direction (\(M_s\) is the saturation magnetization), while \(\Omega\) and \(\partial\Omega\) denote the sample volume and its boundary, respectively.

The surface contribution with anisotropy constant \(K_s > 0\), corresponding to homeotropic anchoring in CLCs or PMA in ChM, incorporates interfacial effects and extends the bulk Dzyaloshinskii model by imposing boundary-induced alignment. {\color{black} This surface contribution is analogous to the Rapini--Papoular anchoring energy widely used in liquid crystals to describe the preferential alignment of the director at confining surfaces~\cite{rapini1969distorsion}. 
In the present magnetic context, the corresponding term imposes homeotropic alignment of the magnetization at the film surfaces and therefore plays a role closely related to surface anchoring in cholesteric liquid crystals.}

For completeness, we also mention a bulk uniaxial anisotropy term,
\begin{equation}
    w_a = -K_u (\mathbf{m} \cdot \mathbf{z})^2.
\label{UA}
\end{equation}
In cholesteric liquid crystals, such a uniaxial anisotropy is typically induced by an applied electric field \(\mathbf{E}\), leading to an energy contribution of the same functional form,
\[
w_a=-\frac{\varepsilon_0 \Delta \varepsilon}{2} (\mathbf{m} \cdot \mathbf{E})^2,
\]
where \(\Delta \varepsilon\) denotes the dielectric anisotropy.

{\color{black}
This expression is, however, strictly valid only in the limit of weak dielectric anisotropy and under the assumption of a spatially uniform electric field. In the general case, the field-induced contribution to the free energy acquires a more complicated form, since the electric field may produce nonlocal electrostatic effects. Such nonlocal contributions have been analyzed, for example, in studies of electric-field-driven cholesteric fingers in Refs.~\cite{shiyanovskii2001director,shiyanovskii2001computer}.
In the present work, we neglect these spatial inhomogeneities of the induced anisotropy, as well as higher-order surface contributions such as the $K_{13}$ and $K_4$ terms~\cite{Selinger,Oswald}. }

\(\mathbf{H}\) the applied magnetic field oriented parallel to the \(z\) axis. {\color{black} In principle, a Zeeman-like term can be substituted by the interplay of an electric field  and the surface anchoring acting in an opposite way. Interestingly, in ferromagnetic CLCs formed by colloidal dispersion of magnetic monodomain nanoparticles, one may achieve a linear coupling and facile response to applied magnetic fields, as well \cite{mertelj2013ferromagnetism,liu2016biaxial}.
Moreover, in recently discovered ferroelectric nematic phases, a spontaneous macroscopic polarization gives rise to a genuine linear coupling to an applied electric field, thereby providing a closer liquid-crystal analogue of a Zeeman-type term~\cite{lavrentovich2020ferroelectric}. }

The bulk energy terms in Eq.~(\ref{functional}) are directly analogous to the elastic contributions of the Frank--Oseen free energy~\cite{Oswald,kleman2003} in the one-constant approximation \(K_1 = K_2 = K_3 = K\)~\cite{kleman2003}. Here, the constants \(K_i\) characterize splay (\(K_1\)), twist (\(K_2\)), and bend (\(K_3\)) distortions of the director field. Within this framework, the material parameters are related via \(A \rightarrow K/2\) and \(D \rightarrow K q_0\), where \(q_0 = 2\pi/\lambda\) is the chiral wave number of the equilibrium cholesteric phase and \(\lambda\) denotes the helicoidal pitch. For many common cholesteric liquid crystals, the elastic constants are of comparable magnitude, rendering the one-constant approximation a reasonable description of their standard elastic behavior.

{\color{black} We emphasize, however, that the minimal model~(\ref{functional}) is not intended to reproduce the full complexity of phase diagrams in cholesteric liquid crystals, which are known to depend sensitively on elastic anisotropy associated with the distinct elastic constants $K_1$, $K_2$, and $K_3$. Rather, the present approach provides a controlled theoretical framework that isolates the essential topological and geometrical mechanisms governing the formation of complex solitonic states and their ordered meta-matter.
We also note that the structure and stability of cholesteric fingers and their lattices beyond the one-constant approximation have been investigated in detail in Ref.~\cite{shiyanovskii2001computer}, where the influence of anisotropic elastic constants was analyzed through numerical simulations and directly compared with experimental observations in cholesteric liquid crystals.}

To render the formulation material-independent, we introduce dimensionless variables. The characteristic length scale arising from the competition between exchange and DMI is defined as \(L_D = A/D\). The magnetic field is expressed in dimensionless form as \(h = \mu_0 M_s H A / D^2\).

We consider a thin-film geometry, infinite in the $x$- and $y$-directions, with a thickness along $z$ equal to one spiral pitch, $\lambda = 4\pi L_D$. 
{\color{black} Such a  geometry is motivated both by the previous theoretical study of hopfion lattices in Ref.~\cite{tai2018} and by the extensive literature on confined cholesteric liquid crystals. 
For analyses of modulated states in thinner and thicker films, see Refs.~\cite{shigenaga2026cholesteric,tai2018}. Certainly, additional effects associated with the film thickness are expected to arise, but their investigation lies beyond the scope of the present manuscript.}

Surface anchoring is conveniently described by the extrapolation length \(\xi = A/K_s\), which quantifies the extent to which the boundary constrains the magnetization. Physically, \(\xi\) corresponds to the hypothetical distance beyond the physical surface at which the magnetization would be perfectly aligned~\cite{tai2018}. The strong-anchoring limit \(\xi \to 0\) enforces a rigid surface orientation.
Scaling lengths by the helix pitch \(\lambda\) leads to a dimensionless extrapolation length \(\overline{\xi} = A / (K_s \lambda)\), which can equivalently be expressed as \(\overline{\xi} = 1/(4\pi k_u^s \overline{t})\), where \(k_u^s\) is the effective uniaxial anisotropy of a narrow surface layer of dimensionless thickness \(\overline{t}\). In this work, we parameterize the surface anchoring using \(k_u^s\) and fix \(\overline{t} = 4\pi/N_z \approx 0.2\), with \(N_z = 64\) being the number of grid points along $z$.
The numerical grid size in the $x$ and $y$ directions is chosen according to the specific simulation: typically, $N_y = 128$, $N_x = 1$ for cholesteric fingers, and $N_x = N_y = 400$ for hopfions.

The complexity of many magnetic  structures does not allow simple analytic descriptions of the director configuration. Therefore, the energy minimization of the functional~(\ref{functional}) is carried out primarily using the GPU-accelerated code \textsc{mumax3} (version~3.10), which integrates the Landau--Lifshitz equation within a finite-difference framework to obtain relaxed magnetization configurations~\cite{mumax3}. 
The reliability of these numerical results is assessed independently by comparison with solutions obtained from in-house implementations of simulated annealing and single-spin Metropolis Monte Carlo algorithms. 
Details of these auxiliary methods have been reported previously~\cite{leonov2021} and are therefore not reiterated here.

We stress that the model~(\ref{functional}) is employed here primarily as a mathematical and phenomenological framework. Any direct comparison of the obtained solutions with experimental realizations in ChM or CLC therefore requires careful consideration. In particular, the strong homeotropic surface anchoring naturally present in CLC is generally difficult to achieve in magnetic systems. Although sizable effective PMA have been reported at interfaces of strained chiral magnets~\cite{karhu2012chiral,huang2012extended}, at magnetic metal--oxide interfaces~\cite{dieny2017perpendicular}, in metallic multilayers~\cite{johnson1996magnetic}, and in chiral magnet--ferromagnet heterostructures~\cite{kawaguchi2016skyrmionic,porter2015manipulation}, their strength typically remains below that characteristic of CLC and is likely insufficient to stabilize the solitonic textures considered here. Conversely, even moderate anisotropy in the Frank elastic constants $K_i$ of CLC can significantly reshape the phase diagram, since stability is controlled by the lowest energy scales. Despite these material-specific differences, the present approach provides a controlled and conceptually transparent basis for treating solitonic phenomena in ChM and CLC within a unified theoretical framework.

%%%%%%%%%%%%%%%%%%%%%%%%%%%%%%%%%%%%%%%%%%%%%%%%%%%%%%%%%%%%%%%%%%%%%%%%%%%%%%%%%
\section{Conical state in thin layers of chiral magnets/chiral liquid crystals }

\subsection{Conical and heliconical states}

The conical phase constitutes a key element underlying the principal features of the hopfion physics addressed in this work. We therefore begin with a detailed characterization of the conical state. In the CLC literature, the conical phase is usually referred to as the \emph{Translationally Invariant Configuration} (TIC) phase~\cite{Oswald}.

The conical spiral is a one–dimensional modulated state 
whose wave vector is oriented along the applied magnetic field [Fig. \ref{fig01}(a)]. 
In this state, the magnetization unit vector $\mathbf{m}$ precesses 
around a fixed axis ($z$) and can be parameterized in spherical coordinates as
\begin{equation}
\mathbf{m}(z) =
\left(
\sin\theta(z)\cos\psi(z),\;
\sin\theta(z)\sin\psi(z),\;
\cos\theta(z)
\right),
\end{equation}
where $\theta(z)$ is the polar angle and $\psi(z)$ is the azimuthal angle. 
For the conical state, both angles depend only on the coordinate $z$.

By contrast, in bulk chiral magnets, the cone angle $\theta$ is spatially uniform 
throughout the sample and deviates only weakly from its equilibrium value, 
for example due to cubic or exchange anisotropies (see, e.g., Ref.~\cite{leonov2026low}). 
Uniaxial anisotropy and an external magnetic field tend to close the cone, 
thereby providing an efficient mechanism for its suppression. 
In bulk ChM, the conical phase is often regarded as the principal competitor of other modulated phases, in particular skyrmion crystals, restricting their stability region to a narrow A-pocket near the ordering temperature~\cite{muhlbauer2009,Crisanti}.

{\color{black}We also note that the conical state has several direct analogs in liquid-crystal physics beyond the classical TIC. 
In particular, closely related heliconical structures arise in twist--bend nematic liquid crystals, theoretically predicted by Meyer and Dozov~\cite{dozov2001spontaneous} and later confirmed experimentally~\cite{borshch2013nematic,chen2013chiral}. 
A similar oblique helicoidal state was also predicted for cholesteric liquid crystals in external fields by Meyer~\cite{meyer1969distortion} and de~Gennes~\cite{de1968calcul}, and subsequently observed experimentally in Ref.~\cite{xiang2014electrooptic}. The influence of homeotropic and planar anchoring on such heliconical cholesteric states has been analyzed in Ref.~\cite{iadlovska2018tuning}. 
More recently, related heliconical structures have also been reported in ferroelectric liquid-crystal systems~\cite{karcz2024spontaneous}. These analogies further emphasize the universal character of conical and heliconical modulations across different classes of chiral soft and magnetic matter.}

\subsection{Analytical approach.}

In the absence of surface anchoring, the conical phase admits an analytical solution of the Euler–Lagrange equations derived from Eq.~(\ref{functional}), given by
\begin{equation}
    \psi_0(z) = -\frac{z}{2}, \qquad \cos \theta_0 = 2h.
    \label{anal}
\end{equation}
This solution describes a single-harmonic helical modulation with a constant cone angle $\theta_0$, while the phase $\psi_0$ varies linearly along the $z$-axis. The point $A$ in the phase diagram [Fig.~\ref{fig01}(b)] marks the termination of the conical state at $\theta_0 = 0$ and corresponds to the critical field value $h = 1/2$.

%\subsection{Energy Functional}

In thin films with the surface anchoring, the one–dimensional bulk energy density  (\ref{functional}) for the conical phase reads as
\begin{equation}
w =
(\theta')^2
+ \sin^2\theta (\psi')^2
- \sin^2\theta \psi'
- k_u \cos^2\theta
- h \cos\theta .
\label{cone}
\end{equation}

Variation of (\ref{cone}) with respect to $\theta$ and including the surface terms gives the following boundary conditions
\begin{equation}
2\theta' + k_s \sin(2\theta) = 0
\quad
\text{at } z=0,\lambda.
\end{equation}
In the limit of strong surface anisotropy, $k_s \gg 1$,
the boundary condition reduces to
\begin{equation}
\sin(2\theta)=0,
\end{equation}
which yields
\begin{equation}
\theta(0)=0,
\qquad
\theta(\lambda)=0.
\end{equation}
Thus, the magnetization is forced to align with the $z$ axis
at both surfaces.

Minimization with respect to $\psi$ yields the following solution similar to bulk ChM in (\ref{anal})
\begin{equation}
\psi(z) = -\frac{z}{2}.
\label{psi}
\end{equation}

The exact profile $\theta(z)$ is obtained from the first integral of the
Euler--Lagrange equation and can be written in implicit form as
\begin{equation}
z =
\int_0^{\theta(z)}
\frac{d\vartheta}{
\sqrt{
C
- \frac{1}{4} \sin^2\vartheta
- k_u \cos^2\vartheta
- h \cos\vartheta
}
}.
\label{zz}
\end{equation}
The integration constant $C$ is determined by the turning-point condition
$\theta' = 0$ at $\theta = \theta_{\max}$, where $\theta_{\max}$ denotes
the maximal cone angle. Substituting this condition into the first integral
yields
\begin{equation}
C =
\frac{1}{4} \sin^2\theta_{\max}
+
k_u \cos^2\theta_{\max}
+
h \cos\theta_{\max}.
\end{equation}
This representation eliminates the integration constant in favor of the
physically meaningful parameter $\theta_{\max}$ and ensures that the
integrand vanishes at $\theta=\theta_{\max}$.

For a layer of thickness $\lambda$, the half-period condition reads
\begin{equation}
\frac{\lambda}{2}
=
\int_{0}^{\theta_{\max}}
\frac{d\vartheta}{\sqrt{F(\vartheta;\theta_{\max})}},
\label{thetamax}
\end{equation}
where $F(\vartheta;\theta_{\max})$ denotes the expression under the square root
in Eq.~(\ref{zz}).

Numerically, this equation can be solved for $\theta_{\max}$ using a
one–dimensional root-finding procedure. For each trial value of $\theta_{\max}$, the integral is evaluated by standard numerical quadrature until convergence of the condition above
is achieved.
Once $\theta_{\max}$ is determined, the spatial profile $\theta(z)$
is reconstructed using (\ref{zz}).
In practice, $z(\theta)$ is computed on a dense grid
$\theta \in [0,\theta_{\max}]$ using numerical quadrature,
and the function $\theta(z)$ is obtained by interpolation.
The second half of the profile follows from symmetry,
$\theta(\lambda - z) = \theta(z)$,
which yields the complete cone configuration across the layer.

%%%%%%%%%%%%%%%%%%%%%%%%%%%%%%%%%%%%%%%%%%%%%%%%%%%%%%%%%%%%%%%%%%%%%%%%%%%%%%%%%%%%%%%%%%%%%%
\subsection{Numerical integration: shooting method}

Since the nonlinear differential equation for the cone angle is written as
\begin{equation}
\theta'' = \sin\theta
\left(
\left(k_u-\frac{1}{4}\right)\cos\theta + \frac{h}{2}
\right),
\label{theta}
\end{equation}
it can be solved numerically using the shooting method, which provides an alternative to the approach described in the preceding section.
 Historically, the shooting method was already employed in early studies of chiral magnetic textures to obtain solutions for isolated skyrmions and skyrmion lattices within the circular-cell approximation \cite{Bogdanov94,Bogdanov99}, long before large-scale micromagnetic simulations based on finite-difference schemes such as \texttt{mumax3} \cite{mumax3} became standard.

In this approach, the boundary-value problem is reduced to an initial-value problem by treating the initial slope as an adjustable parameter and iteratively tuning it until the boundary conditions at the opposite surface are satisfied.

First, we convert Eq. (\ref{theta}) into a system of first–order ordinary differential equations.  Introducing
\begin{equation}
y_1 = \theta,
\qquad
y_2 = \theta',
\end{equation}
we obtain
\begin{equation}
\label{system}
\left\{
\begin{aligned}
y_1' &= y_2, \\
y_2' &= \sin y_1
\left(
\left(k_u-\frac{1}{4}\right)\cos y_1 + \frac{h}{2}
\right).
\end{aligned}
\right.
\end{equation}

%The boundary conditions are imposed at the two ends of the layer, $\theta(0)=\theta(\lambda)=0$.
%
At $z=0$, we fix
\begin{equation}
y_1(0)=0,
\end{equation}
and treat the initial slope
\begin{equation}
y_2(0)=\theta'(0)
\end{equation}
as an adjustable parameter.

Then, for a given trial value of $y_2(0)$, the system (\ref{system})
is integrated numerically from $z=0$ to $z=\lambda$,
for example using a fourth–order Runge–Kutta scheme or any standard
ODE solver.

After integration, the terminal value $y_1(\lambda)$ is evaluated.
The correct physical solution must satisfy $\theta(\lambda)=y_1(\lambda)=0$. 

The initial slope $y_2(0)$ is iteratively adjusted (e.g.\ by a
bisection, secant, or Newton procedure) until the shooting condition
is fulfilled within a prescribed numerical tolerance.

This procedure selects the unique trajectory in phase space that
satisfies both boundary conditions and yields the true conical
spiral profile $\theta(z)$ in the finite layer.

\subsection{Failure of linearization}

One might attempt to linearize the equation (\ref{theta}) for small cone angles by
using the approximations
\begin{equation}
\sin\theta \approx \theta,
\qquad
\cos\theta \approx 1.
\end{equation}
This is a standard approach when treating nonlinear differential equations
in the small-amplitude limit.

With these assumptions, the governing nonlinear equation immediately reduces
to the linear form
\begin{equation}
\theta'' + \left(k_u + \frac{h}{2} - \frac{1}{4}\right)\theta = 0.
\end{equation}

The corresponding solution is
\begin{equation}
\theta(z) \sim \sin(\kappa z),
\qquad
\kappa^2 = k_u + \frac{h}{2} - \frac{1}{4}.
\end{equation}

However, in the presence of surface anisotropy, the boundary conditions
require $\theta(0)=\theta(\lambda)=0$.
For arbitrary values of $k_u$, $h$, and $\lambda$, the condition
$\kappa \lambda = n\pi$ (with integer $n$) is generally not satisfied.
Therefore, a simple sinusoidal profile cannot simultaneously fulfill
both boundary conditions.

The linearized theory thus fails to describe the conical state in finite
layers. In particular, it neglects the pronounced suppression of
$\theta(z)$ near the surfaces imposed by surface anisotropy.
As a consequence, even for small maximal angles $\theta_{\max}$,
the cone profile remains intrinsically nonlinear.
The nonlinearity originates from the boundary pinning term
proportional to $k_s$, which cannot be captured within
the small-angle approximation.

%%%%%%%%%%%%%%%%%%%%%%%%%%%%%%%%%%%%%%%%%%%%%%%%%%%%%%%%%%%%%%%%%%%%%%%%%%%%%%%%%%%%%%%%%%%
\subsection{Stability and Saturation of the Confined Conical Phase}

Despite the valuable physical insight gained from the analytical solutions presented in the preceding sections, general solutions for the confined conical state are most conveniently obtained using \texttt{mumax3} and are summarized in Fig.~\ref{fig01}. 

According to the phase diagram in Fig.~\ref{fig01}(b), the conical phase saturates along the line $A$--$B$. Interestingly, this second-order phase transition line initially decreases sharply with increasing surface anchoring strength, but quickly approaches a plateau, reaching an almost constant critical field value even for very strong surface anchoring.

Figure~\ref{fig01}(c) summarizes color maps of the energy density distribution for the individual energy contributions in Eq.~(\ref{functional}) at $h = 0$ and $k_u^s = 20$. The DMI energy density (first panel in Fig.~\ref{fig01}(c)) is negative in the middle of the layer and decreases in magnitude toward the surfaces, where the magnetization rotation is suppressed. The exchange energy (second panel in Fig.~\ref{fig01}(c)), by contrast, attains its largest positive value in the center of the film, while it becomes smaller near the surfaces, where the spins tend to align more uniformly. The total energy density (third panel in Fig.~\ref{fig01}(c)) closely follows the behavior of the DMI contribution, reflecting its dominant role in stabilizing the conical state.

In Fig.~\ref{fig01}(d), we plot the magnetization profile across the film thickness. The magnetization, being nearly aligned along the $z$ axis at the surfaces, reaches its minimal value of $m_z$ in the center of the layer. Figure~\ref{fig01}(e) summarizes the dependence of $m_{z,\min}$ on the applied field, up to the point where it reaches unity, signaling the transition to the homogeneous state. The value of $m_{z,\min}$ corresponds to $\theta_{\max} = \arccos(m_{z,\min})$ used in Eq.~(\ref{thetamax}).

\begin{figure*}
  \centering
  \includegraphics[width=0.99\linewidth]{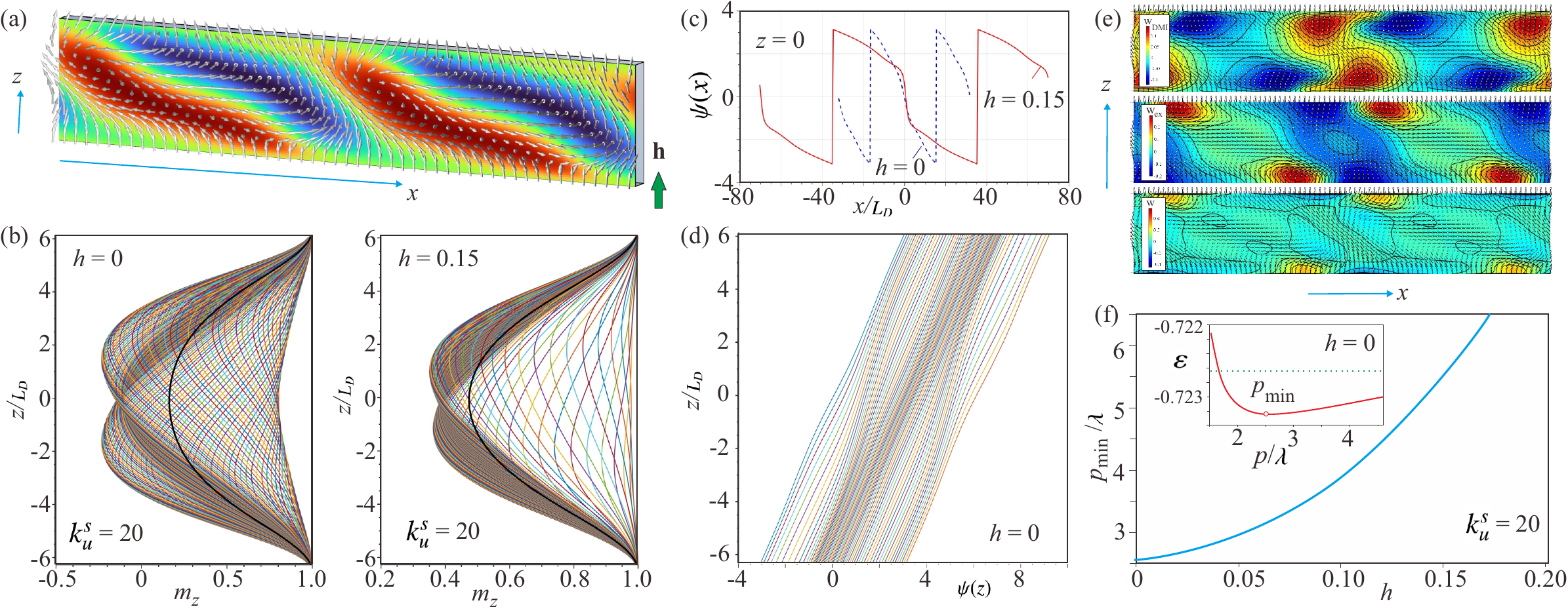}
\caption{\label{fig02}
(Color online) Internal structure and energetic properties of the CF--1 phase.
(a) Representative magnetization configuration of a cholesteric finger of the first type.
(b) Angular profiles $\theta(z)$ taken at different lateral positions $x$ across one period of the CF--1 structure. The black solid line shows the profile for the conical phase for comparison. Most CF--1 profiles exhibit stronger rotation, with some reaching negative values of $m_z$, while profiles with reduced rotation compensate this excess. With increasing magnetic field ($h=0.15$, $k_u^s=20$, second panel), the profiles progressively converge toward the conical solution.
(c) Lateral dependence $\psi(x)$ at the mid-plane $z=\lambda/2$, revealing kink-like behavior reminiscent of domain wall solutions of the sine--Gordon model.
(d) Corresponding profiles of the azimuthal angle $\psi(z)$ demonstrating deviations from the strictly linear dependence characteristic of the conical phase.
(e) Spatial distribution of the energy density contributions calculated relative to the conical phase, highlighting regions of energetic gain and loss within the CF--1 structure.
(f) Inset: Dependence of the CF--1 energy density $\varepsilon$ (\ref{epsilon}) on the finger period $p$ showing a clear minimum that determines the equilibrium period $p_{min}$. Main panel: equilibrium period $p_{\mathrm{min}}$ as a function of the applied magnetic field, demonstrating gradual expansion of the CF--1 structure.}
\end{figure*}

%%%%%%%%%%%%%%%%%%%%%%%%%%%%%%%%%%%%%%%%%%%%%%%%%%%%%%%%%%%%%%%%%%%%%%%%%%%%%%%%%%%%%%%%%%%%
\section{Internal structure of cholesteric fingers CF--1 and CF--2}
\label{sect:structure}
%%%%%%%%%%%%%%%%%%%%%%%%%%%%%%%%%%%%%%%%%%%%%%%%%%%%%%%%%%%%%%%%%%%%%%%%%%%%%%%%%%%%%%%%%%%%

Cholesteric fingers of the first and second type are among the earliest and most extensively studied localized modulated structures in chiral liquid crystals. 
{\color{black} The terminology CF--1 and CF--2 adopted throughout the present manuscript follows the classification introduced in the pioneering works of Oswald and co-workers~\cite{oswald2000static,baudry1999looped}. Detailed theoretical and experimental investigations of the internal structure and stability of cholesteric fingers of the first and second type were subsequently reported in Ref.~\cite{smalyukh2005electric}. A modern perspective on the topology, energetics, and interactions of cholesteric fingers from a magnetic viewpoint is provided in Ref.~\cite{shigenaga2026cholesteric}.}
Experimentally, cholesteric fingers appear as elongated, stripe-like textures embedded in an otherwise homogeneous or weakly twisted background (see, e.g., Refs.~\cite{meyer1976,de1995,Oswald}). To the best of our knowledge, the first numerical investigations of CF--1 date back to the 1970s and can be traced to Ref.~\cite{press1976static}. Numerical solutions for cholesteric fingers of the second type were also obtained at an early stage within continuum models employing the simplifying assumption of isotropic elasticity; see, for example, Ref.~\cite{gil1998surprising}. Subsequent experimental studies highlighted their remarkable robustness under applied electric fields and the associated nonlinear electro-optical response~\cite{ribiere1994optical}.

Accordingly, our aim is not to reproduce or refine this extensive body of work, but rather to place cholesteric fingers into the context of hopfions, which can be viewed as circular realizations of the corresponding finger textures [see Ref. \cite{shigenaga2026cholesteric}]. 
Moreover, recent work~\cite{hopfions} has suggested that the metastability region of hopfions is intimately connected to the stability domains of the corresponding cholesteric fingers, implying that hopfions may be regarded as precursor states of these finger configurations. 
Therefore, in the following we provide a refined description of the internal structure and the field-driven evolution of CF--1 and CF--2.

%%%%%%%%%%%%%%%%%%%%%%%%%%%%%%%%%%%%%%%%%%%%%%%%%%%%%%%%%%%%%%%%%%%%%%%%%%%%%%%%%%%%%%%%%%
\subsection{Stability of the modulated CF-1 phase \label{sect:CF1stability}}
%%%%%%%%%%%%%%%%%%%%%%%%%%%%%%%%%%%%%%%%%%%%%%%%%%%%%%%%%%%%%%%%%%%%%%%%%%%%%%%%%%%%%%%%%%%

The internal structure of cholesteric fingers of the first type [Fig.~\ref{fig02}(a)] can be understood on the basis of the conical state analyzed in the preceding section. Within CF-1, the angular profiles $\theta(z)$ exhibit the same qualitative behavior: the spins rotate from the state $\theta(0)=0$ at the lower surface, reach a maximal value $\theta_{\max}$, and then rotate back to $\theta(\lambda)=0$ at the upper surface. The crucial difference from the conical state is that $\theta_{\max}$ is reached not at the middle of the layer, but at different heights depending on the lateral coordinate $x$, i.e., in the CF-1 state $\theta=\theta(x,z)$. 

Figure~\ref{fig02}(b) (first panel) shows the angular profiles $\theta(z)$ across one period of the finger phase for $h=0, k_u^s=20$. Interestingly, the majority of these profiles exhibit a rotation exceeding that of the conical phase (shown by the black solid curve): the corresponding values of $m_{z,\min}$ may even become negative. The gain in rotational energy is compensated by other profiles with reduced rotational amplitude. At the same time, no angular profile with a strictly homogeneous distribution $\theta(z)=0$ was found.
With increasing magnetic field, more profiles within the CF-1 phase converge toward that of the conical phase. Correspondingly, the density of rotationally impeded profiles becomes scarce [Fig.~\ref{fig02}(b), second panel, $h=0.15, k_u^s=20$].

The azimuthal angle $\psi$ also becomes a function of both spatial coordinates, $\psi=\psi(x,z)$. The lateral profiles $\psi(x)$ taken at the mid-plane $z=\lambda/2$ exhibit a pronounced nonlinear character and closely resemble domain wall profiles or kink-like solutions known from the sine--Gordon theory [Fig.~\ref{fig02}(c)].
Along the thickness direction $z$, the dependence of $\psi$ deviates from the strictly linear behavior characteristic of the conical phase (\ref{psi}), although these deviations remain relatively weak [Fig.~\ref{fig02}(d)]. 

The characteristic energy density distributions [shown in the same sequence as for the conical phase in Fig.~\ref{fig01}(c)] are presented relative to the conical state [Fig. \ref{fig02}(e)]; specifically, at each point $(x,z)$ the local energy density of the conical phase is subtracted. This representation highlights the regions of the finger structure that are energetically favorable compared to the cone. All energy fingerprints display alternating positive and negative contributions, most prominently near the surfaces. Notably, the rotational DMI energy within the CF-1 phase attains a slightly higher magnitude than in the conical state. In contrast, the exchange energy is reduced, and it is precisely this reduction that provides the overall energetic advantage of the CF-1 phase. In Fig.~\ref{fig01}(b), the curvilinear region $a$--$b$--$c$ marks the parameter domain where the CF-1 phase possesses lower energy than the conical phase.

We emphasize  that the energy density of the CF--1 phase, defined as 
\begin{equation}
    \varepsilon = \frac{1}{p\lambda} \int w(x,z)\, dxdz,
    \label{epsilon}
\end{equation}
 is minimized with respect to the period $p$ at every point of the phase diagram in Fig.~\ref{fig01}(b). The inset in Fig.~\ref{fig02}(f) shows a representative dependence $\varepsilon(p)$ for $h=0$ and $k_u^s=20$, exhibiting a clear minimum that defines the equilibrium period. The resulting values of $p_{\mathrm{min}}$, plotted in Fig.~\ref{fig02}(f) as a function of the magnetic field, reveal a gradual expansion of the CF--1 structure. 

The equilibrium period substantially exceeds the film thickness~$\lambda$, reflecting the strong lateral expansion of the CF--1 structure.
Nevertheless, the bulk uniaxial anisotropy (\ref{UA}) plays a decisive role in preserving the localized nature of the fingers [this mechanism will be analyzed elsewhere \cite{shigenaga2026cholesteric}]. By imposing an energetic cost on transverse deviations of the magnetization, it effectively confines the twisted region and prevents its unlimited spreading. In the absence of bulk uniaxial anisotropy, CF--1 loses this confinement: the fingers progressively broaden, and their period tends toward divergence, resulting in a delocalized, weakly modulated state. Therefore, the bulk uniaxial anisotropy (\ref{UA}) is the key factor that stabilizes CF--1 as a localized object with a finite spatial extent. In the following section, we demonstrate that CF--2 remains substantially more compact even within the same model.

To the best of our knowledge, no experimental study of the CF--1 phase has reported such a continuous increase of its period up to divergence. This is in sharp contrast to the classical experimental realization of the Dzyaloshinskii scenario in chiral magnets without surface anisotropy, where the chiral soliton lattice expands with increasing magnetic field by releasing isolated kinks, leading to a divergence of the period, as demonstrated in Ref.~\cite{togawa2012chiral}.

\begin{figure*}
  \centering
  \includegraphics[width=0.99\linewidth]{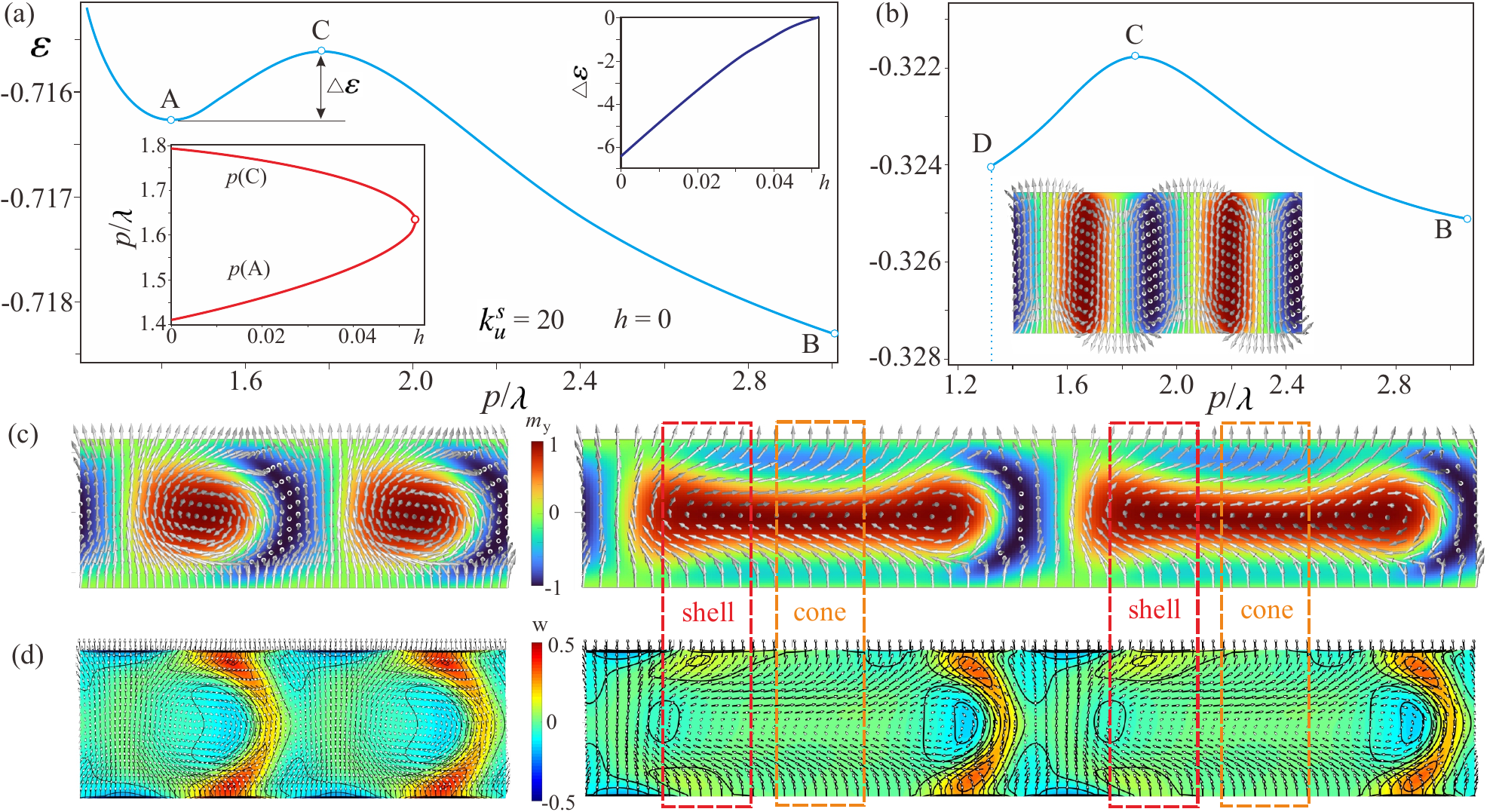}
\caption{\label{fig03}
(a) Energy density of the bimeron lattice as a function of the lattice period $p$ at fixed control parameters $k_u^s=20$ and $h=0$, showing a local minimum ($A$) separated by a potential barrier (maximum at $C$) from the lower-energy state with isolated bimerons embedded in the conical phase ($B$). Left inset: magnetic-field dependence of the lattice periods corresponding to points $A$ and $C$ up to the critical field where the minimum and maximum merge. 
Right inset: field dependence of the potential barrier height $\Delta \varepsilon$, which vanishes at the critical field.
(b) Energy-density curve in the regime where the minimum disappears from the left side, leading to a discontinuous transition of the bimeron lattice into the energetically favorable spiral state (spin distribution shown in the inset). 
(c),(d) Spatial distribution of the local energy density $w(x,z)$ (calculated relative to the energy density of the conical state) for representative lattice periods corresponding to points $A$ and $B$. The shell regions with positive energy density are indicated by dashed red rectangles, while the domains of the penetrating conical phase are highlighted by dashed orange rectangles. 
}
\end{figure*}

 %%%%%%%%%%%%%%%%%%%%%%%%%%%%%%%%%%%%%%%%%%%%%%%%%%%%%%%%%%%%%%%%%%%%%%%%%%%%%%%%%%%%%%%%%%
\subsection{Stability of the modulated CF--2/bimeron phase \label{sect:CF2stability}}
%%%%%%%%%%%%%%%%%%%%%%%%%%%%%%%%%%%%%%%%%%%%%%%%%%%%%%%%%%%%%%%%%%%%%%%%%%%%%%%%%%%%%%%%%%%

The composite objects shown in Fig.~\ref{fig03}(c) are commonly referred to as \emph{bimerons} in the skyrmion literature~\cite{lin2015,ohara2022}. In CLC, the same solitonic structures are known as cholesteric fingers of the second type (CF--2)~\cite{Oswald}. In the following, we use both terms interchangeably to refer to this soliton.

Our analysis of bimeron stability starts from the point $h = 0$, $k_u^s = 20$ in the phase diagram shown in Fig.~\ref{fig01}(b), and proceeds by minimizing the energy density $\varepsilon(p)$ (\ref{epsilon}) of the modulated bimeron phase with respect to its period, as illustrated in Fig.~\ref{fig03}(a). This dependence exhibits noticeable deviations from the general behavior exemplified by CF--1 in Fig.~\ref{fig02}(f) [inset].

First, the energy curve $\varepsilon(p)$ displays a well-defined minimum corresponding to an equilibrium inter-bimeron separation $p_{\min}$ (point $A$ in Fig.~\ref{fig03}(a)). The spin configuration at this point is shown in Fig.~\ref{fig03}(c),(d) [left panels] as color maps of the $m_y$ component of the magnetization and of the total energy density $w(x,z)$. The equilibrium state, however, is separated from the conical state by a finite energy barrier $\Delta \varepsilon$ (point $C$ in Fig.~\ref{fig03}(a)), indicating that the bimeron phase is metastable and possesses a higher energy than the conical phase.

Upon overcoming this energy barrier, the distance between neighboring bimerons increases continuously and without bound, allowing the conical phase to penetrate into the inter-bimeron regions. The corresponding spin configuration at point $B$ in Fig.~\ref{fig03}(a) is shown in Fig.~\ref{fig03}(c),(d) [right panels]. The domains of the conical spiral are highlighted by dashed orange rectangles.

%The origin of the potential barrier in Fig. \ref{fig03}(a) becomes evident from the energy density distribution $w(x,z)$ shown in Fig.~\ref{fig03}(d) [bottom right panel]. As the system evolves from point $A$ toward point $C$, the energy density $\varepsilon(p)$ of the bimeron phase increases until the conical state begins to penetrate and forms an energetically costly region (referred to as a ``shell'') around each bimeron. At this stage, the energy density reaches its maximal value, $\varepsilon(C)$. Once the shell is fully formed, the energy density decreases toward point $B$, as the fraction of conical domains with lower energy density progressively increases. The shell region characterized by a positive energy density $w(x,z)$ is indicated by dashed red rectangles in Fig.~\ref{fig03}(c),(d).

The origin of the potential barrier in Fig.~\ref{fig03}(a) can be understood as a competition between two energy contributions. On the one hand, the conical state possesses a lower energy density than the bimeron cores and therefore lowers the total energy density as its relative fraction within one lattice period increases. On the other hand, the penetration of the conical phase between neighboring bimerons requires the formation of transitional regions at their boundaries, referred to as ``shells'', which carry an energetic penalty.
As the lattice period increases, the shell regions gradually develop and their total energy cost grows, reaching a maximum at point~$C$. At this stage, the shells are fully formed and the associated interface energy is maximal. However, with further increase of the period, the fraction of the energetically favorable conical phase continues to increase. Beyond a characteristic period $p(C)$, the energy gain due to the growing conical domains outweighs the shell penalty, and the total energy density decreases rapidly toward point~$B$.
The shell regions, characterized by a positive energy density $w(x,z)$, are indicated by dashed red rectangles in Fig.~\ref{fig03}(c),(d).

With increasing magnetic field, the energy-density minimum (and hence the associated potential barrier) progressively disappears, signaling the loss of stability of the bimeron lattice. In this regime, each incremental increase of the lattice period leads to an immediate energy gain, since the penetrating conical domains reduce the total energy more strongly than the shell regions increase it. As a result, the energy-density curve becomes monotonic and no stationary lattice period exists.
The left inset of Fig.~\ref{fig03}(a) presents the field dependence of the bimeron periods corresponding to points~$A$ and~$C$, up to the field at which the minimum and maximum of the energy-density curve merge. The right inset shows the corresponding magnitude of the potential barrier $\Delta \varepsilon$, which vanishes at the critical field.

The metastability region of the bimeron phase is indicated by the curvilinear domain $a$--$d$--$e$ in the phase diagram shown in Fig.~\ref{fig01}(b). Along the upper boundary $d$--$e$, the energy-density minimum corresponding to the equilibrium inter-bimeron separation vanishes, as discussed above, so that a periodic bimeron lattice can no longer be sustained and only isolated bimerons embedded in the conical phase remain stable.
Along the boundary $a$--$d$, in contrast, the energy minimum vanishes from the left side, as illustrated in Fig.~\ref{fig03}(b). In this case, the bimeron lattice undergoes a discontinuous transition into the energetically more favorable spiral state (the spin distribution of the spiral state is shown in the inset of Fig.~\ref{fig03}(b)).

Remarkably, even above the critical field at which the bimeron lattice loses its stability, isolated bimerons embedded in the conical phase continue to attract each other and can therefore form bound clusters. Before analyzing the bimeron--bimeron interaction in detail, we first examine the internal structure of isolated bimerons submerged in the conical phase.

\begin{figure}
  \centering
  \includegraphics[width=0.99\linewidth]{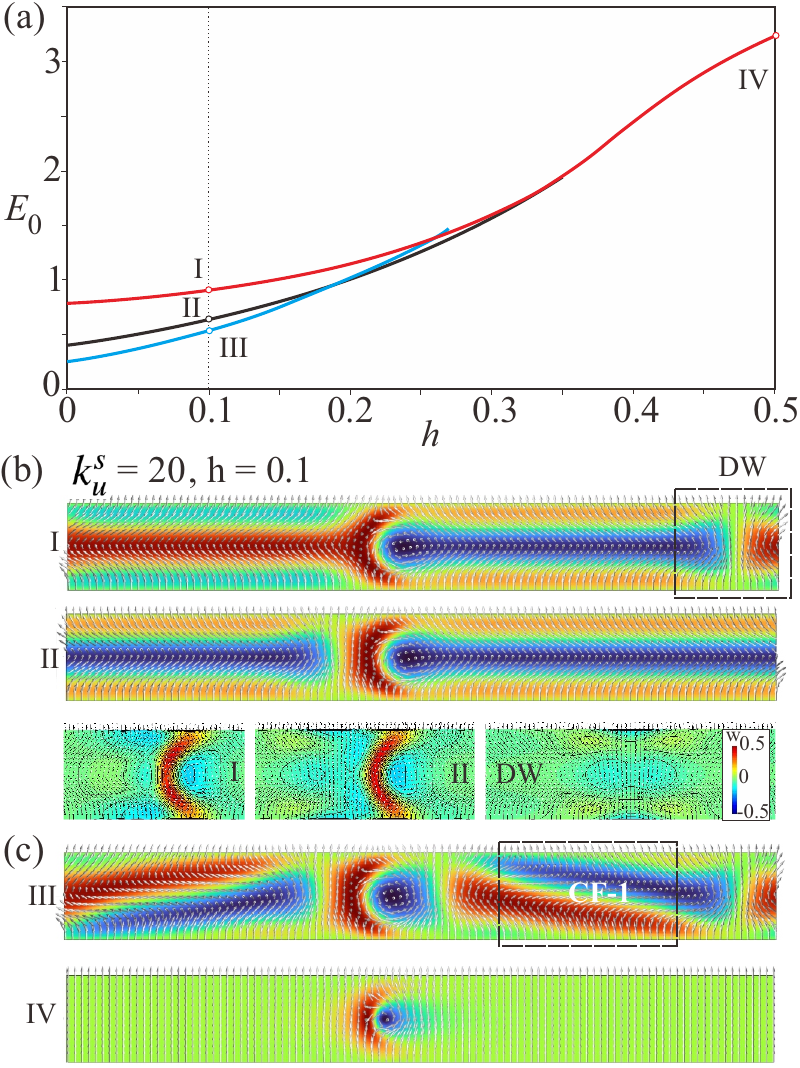}
\caption{\label{fig04}
(a) Eigen-energies of isolated bimerons embedded in the conical phase, calculated relative to the energy of the conical state and plotted as a function of magnetic field. The red and black curves correspond to configurations~I and~II, respectively.  
(b) Spin configurations and corresponding energy-density distributions. Upper row: configuration~I, in which the bimeron induces a phase shift of the conical state and generates a domain wall between conical domains of opposite phase (highlighted by the dashed black rectangle). Second row: configuration~II, where the conical phase has identical phase on both sides of the bimeron and no DW is formed. Third row: spatial distributions of the local energy density for configurations~I and~II, as well as for the isolated DW.  
(c) Spin configuration in which the bimeron is surrounded by the CF--1 phase prompted by domain walls separating conical domains of different phase. At the saturation field corresponding to the line $A$--$B$ in Fig.~\ref{fig01}(b), the bimeron is embedded in the homogeneous state (configuration~IV, lower panel).
}
\end{figure}

\begin{figure*}
  \centering
  \includegraphics[width=0.9\linewidth]{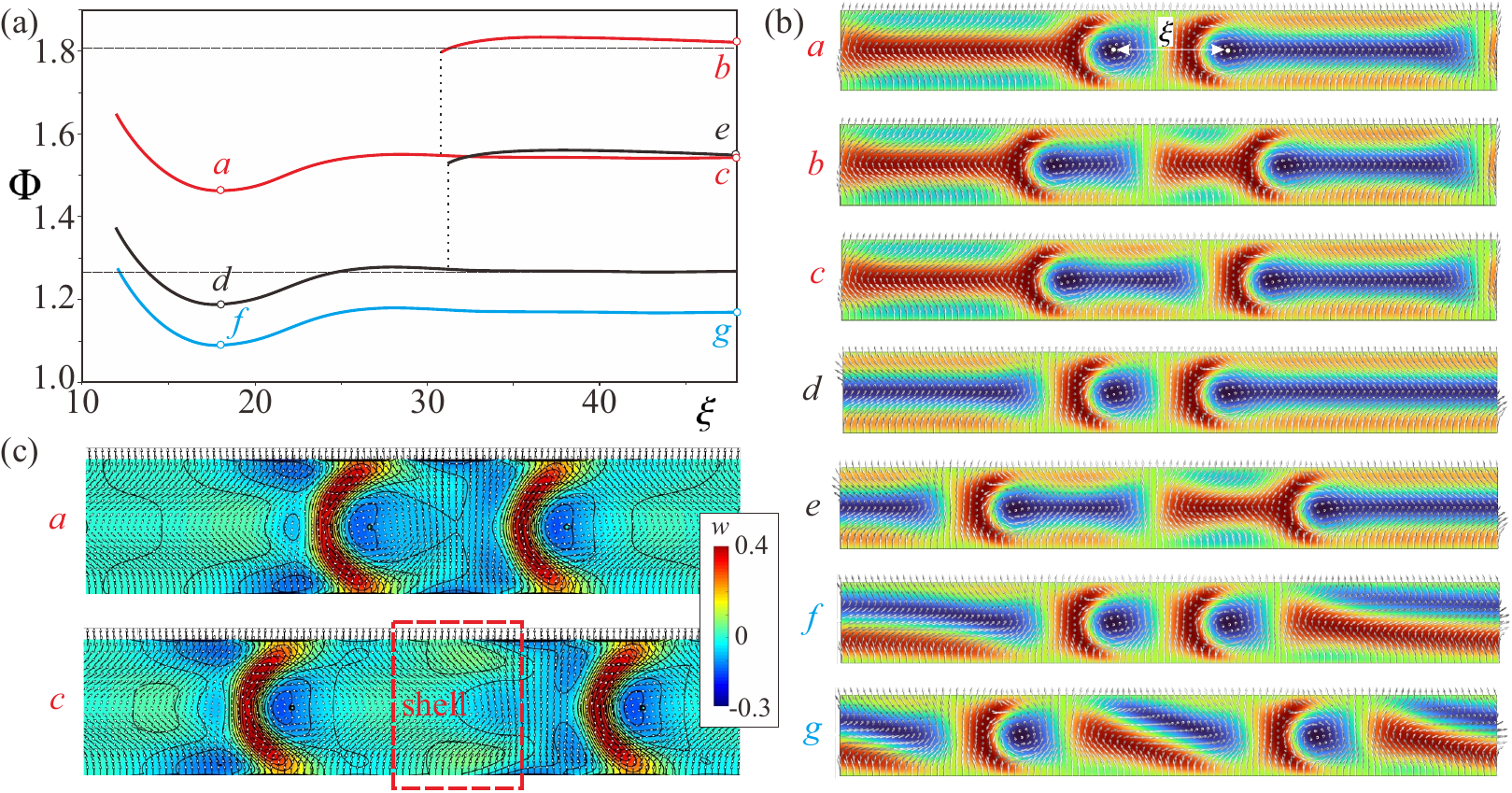}
\caption{\label{fig05}
(a) Interaction potentials $\Phi(\xi)$ for pairs of CF--2 (bimerons) embedded in the conical phase at $k_u^s=20$ and $h=0.1$. The total energy relative to the conical state is plotted as a function of the inter-bimeron separation~$\xi$ for different realizations distinguished by the number and arrangement of domain walls. Branching of the curves reflects the formation or annihilation of DWs upon varying~$\xi$.  
(b) Representative spin configurations corresponding to the labeled states ($a$--$e$). Configuration~$b$ contains two DWs and exhibits the highest pair energy at large separations, while configurations~$c$ and~$e$ contain a single DW located either in the interior or exterior region. Configurations~$a, d, f$ correspond to the energetically most favorable bound states without interior DWs.  
(c) Spatial distributions of the local energy density for selected two-bimeron configurations. The shell region is indicated by the dashed red rectangle. As in the lattice case, the shell carries positive excess energy and consists of surface regions with energy density above that of the conical phase and a central rounded region with negative energy density.
}
\end{figure*}

%%%%%%%%%%%%%%%%%%%%%%%%%%%%%%%%%%%%%%%%%%%%%%%%%%%%%%%%%%%%%%%%%%%%%%%%%%%%%%%%%%%%%%%%%%%%%
\subsubsection{Internal structure of isolated bimerons within the conical state}
%%%%%%%%%%%%%%%%%%%%%%%%%%%%%%%%%%%%%%%%%%%%%%%%%%%%%%%%%%%%%%%%%%%%%%%%%%%%%%%%%%%%%%%%%%%%%

Placement of an isolated bimeron into the conical state permits two distinct realizations, governed by different energetic mechanisms. The corresponding eigen-energies of these spin configurations, calculated relative to the energy of the conical state and plotted as a function of magnetic field, are shown in Fig.~\ref{fig04}(a).

In the first, energetically less unfavorable scenario (red curve in Fig.~\ref{fig04}(a)), the bimeron induces a phase shift of the surrounding conical state between its left and right sides (upper configuration~I in Fig.~\ref{fig04}(b)). The color plot of the $m_y$ component of the magnetization reveals that the blue bimeron core merges with the conical domain on the right, while the red bimeron crescent connects to a conical domain with opposite phase on the left. As a result, a domain wall (DW) forms between the two conical domains (highlighted by the dashed black rectangle), which is energetically favorable in this configuration. 
The eigen-energy of the isolated bimeron is $E_0 = 0.9305$, whereas the energy contribution of the DW is $E_{\mathrm{DW}} = -0.0366$. Structurally, this is the same type of DW that appears within the CF--1 state [Fig.~\ref{fig02}(a)], which therefore complements the description of its internal structure: CF--1 can be viewed as conical domains with different phases separated by domain boundaries possessing negative eigen-energy.

In the second case (black curve in Fig.~\ref{fig04}(a)), the phase of the conical spiral remains identical on both sides of the isolated bimeron (second configuration~II in Fig.~\ref{fig04}(b)). In this realization, no domain wall is formed in the surrounding conical state. The corresponding eigen-energy of the bimeron is $E_0 = 0.6323$. 
The energy-density distributions within configurations~I and~II, as well as within the domain wall, are shown in the third row of Fig.~\ref{fig04}(b).

The formation of energetically favored domain walls may induce the transformation of the surrounding conical phase into the structurally modified CF-1 phase [blue curve in Fig. \ref{fig04}(a) and the spin configuration III in Fig. \ref{fig04}(c), upper panel]. This realization possesses the lowest eigen-energy in some field range until the conical phase returns its energetic advantage [compare with the line $d-e$ at the phase diagram in Fig. \ref{fig01}(b)]. 

At the saturation field corresponding to the line $A$--$B$ [Fig.~\ref{fig01}(b)], the isolated bimeron is embedded in the homogeneous state (spin configuration~IV in Fig.~\ref{fig04}(c), lower panel).

{\color{black} We note, however, that the whole range of bimeron metastability requires extended calculations and the analysis of the collapse mechanism. The full analysis will be done elsewhere \cite{shigenaga2026cholesteric}.}

%%%%%%%%%%%%%%%%%%%%%%%%%%%%%%%%%%%%%%%%%%%%%%%%%%%%%%%%%%%%%%%%%%%%%%%%%%%%%%%%%%%%%%%%%%%%%
\subsubsection{Origin and structure of the attractive inter-bimeron potential}%%%%%%%%%%%%%%%%%%%%%%%%%%%%%%%%%%%%%%%%%%%%%%%%%%%%%%%%%%%%%%%%%%%%%%%%%%%%%%%%%%%%%%%%%%%%%

To quantify the CF2--CF2 interaction within the conical phase, we compute the interaction potentials $\Phi$ for bimeron pairs, shown in Fig.~\ref{fig05}(a), at the parameter point $k_u^s = 20$ and $h = 0.1$, i.e., in the regime where the bimeron lattice is already unstable.
The interaction potential is obtained by evaluating the total energy~(\ref{functional}) for configurations in which the centers of two bimerons are pinned at a prescribed separation~$\xi$.
For clarity, the reference energy of two infinitely separated bimerons is not subtracted; instead, the full total energy relative to the conical state is plotted as a function of~$\xi$.

Different realizations of bimerons within the conical state give rise to different energy ranks of the interaction potential (Fig.~\ref{fig05}(a)). At large separation distances between two bimerons, the pair energy depends on the number of domain walls  formed between conical domains of different phase. The highest interaction energy is obtained for configuration~$b$ [Fig.~\ref{fig05}(b)], which contains two DWs. Configurations~$e$ and~$c$ contain only a single DW, located either in the interior region between the two bimerons or in the exterior region surrounding them.

For large separations, the bimeron pairs are separated by a small energy barrier from the most energetically favorable configuration corresponding to coupled bimerons. The minimum of the interaction potential occurs at the same inter-bimeron distance, irrespective of the number of domain walls present and/or the surrounding phase (conical or CF--1).

Frustration of the conical phase and the associated formation or annihilation of DWs underlie the branching of the interaction potential. As an instructive example, consider two distant bimerons in configuration~$b$ (Fig.~\ref{fig05}(b)). Upon approaching each other, the bimerons eliminate the DW between them and transform into configuration~$a$, corresponding to the minimum of the interaction potential. However, when increasing the separation starting from configuration~$a$, the system follows the red curve in Fig.~\ref{fig05}(a) toward configuration~$c$, which has a lower energy among the remote bimeron pairs. 
A similar hysteretic behavior can be traced between configurations~$e$ and~$d$, as well as toward the energetically lower configuration without an interior DW (not shown). The presence of an intervening CF--1 phase does not qualitatively modify the character of the interaction potential, although it reduces the overall energy.

The formation of the minimum in the interaction potential can be understood on the basis of the shell region introduced above. Figure~\ref{fig05}(c) shows color plots of the energy density for two representative bimeron configurations. The shell region is highlighted by the dashed red rectangle. As in the case of the bimeron lattice, the shell carries a positive excess energy. It consists of two parts: surface regions with energy density higher than that of the conical phase, and a central rounded region characterized by negative energy density. 

At first glance, the absence of a stable bimeron lattice at $h=0.1$ may seem to contradict the existence of an attractive inter-bimeron potential and the tendency of isolated bimerons to form extended clusters. However, these two findings are fully consistent. The stability of a periodic lattice requires the presence of a local minimum of the energy density (\ref{epsilon}) with respect to the lattice period, ensuring a stationary equilibrium separation in the thermodynamic limit. In contrast, the attractive interaction between isolated bimerons reflects only the pairwise energetics at finite separation and does not guarantee the existence of a bulk phase with a well-defined equilibrium period. In the present case, although the pair potential exhibits a pronounced minimum favoring bound states and cluster formation, the total energy density decreases monotonically with increasing period once the conical phase becomes energetically dominant. Consequently, extended assemblies of bimerons should be understood as bound clusters stabilized by attractive interactions rather than as a thermodynamically stable bulk lattice phase.

\begin{figure*}
  \centering
  \includegraphics[width=0.99\linewidth]{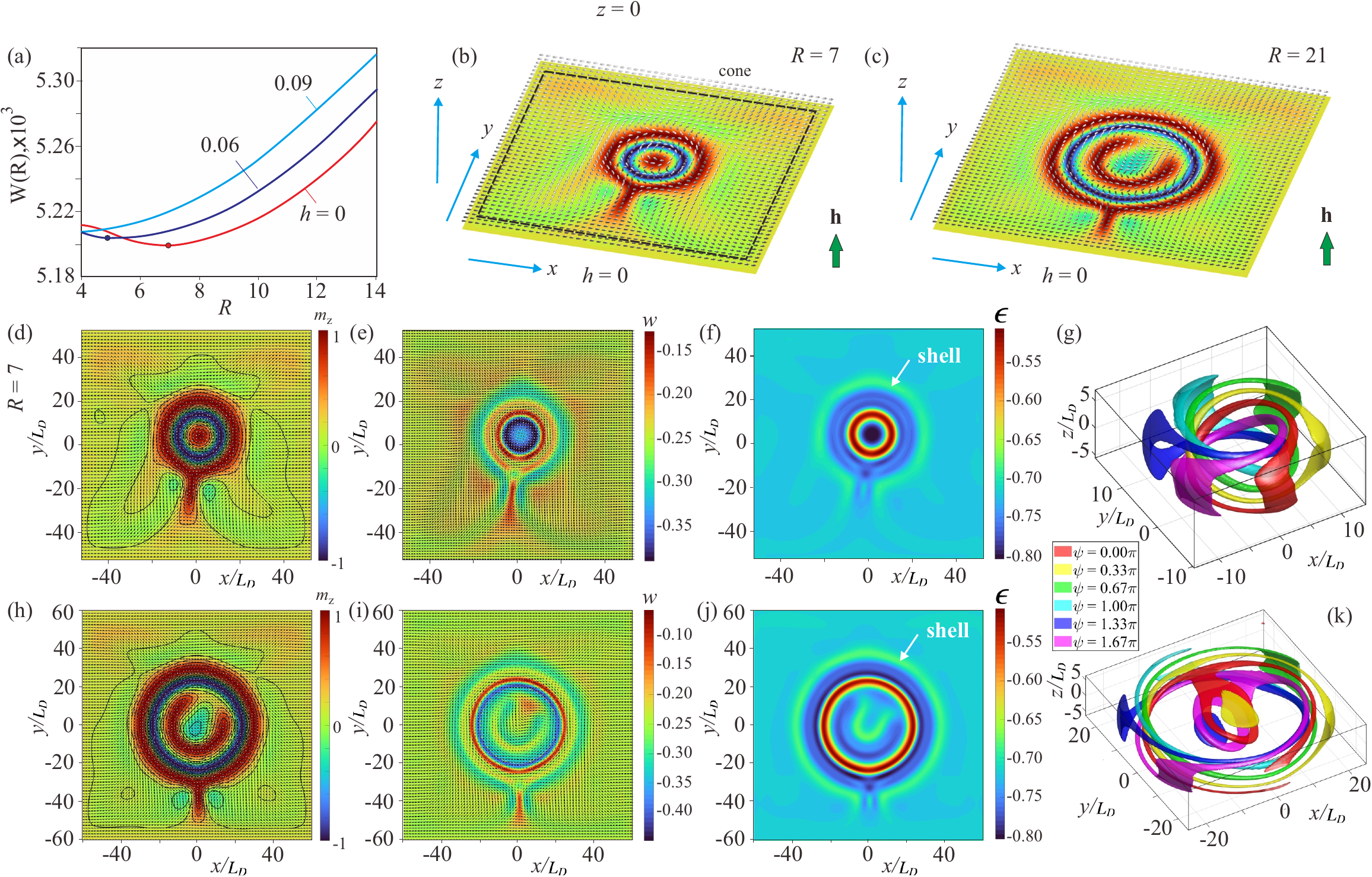}
\caption{\label{fig06} (a) Eigen-energy $W(R)$ of a hopfion as a function of its radius $R$ for several magnetic field values at $k_u^s=20$. The energy is calculated with respect to the conical state. The minimum of $W(R)$ defines the equilibrium hopfion radius, whereas the absence of a minimum indicates instability. 
(b)–(d),(h) Color plots of the $m_z$ component of the magnetization in the middle cross-section for selected points along the energy curve at $h=0$. Panels (b),(c) illustrate the three-dimensional structure of the rotating magnetization, while (d),(h) show the corresponding projections onto the $xy$ plane. The hopfion profile exhibits a characteristic stick-like segment allowing adaptation to the rotating conical spiral. 
(e),(i) Energy density distributions $w(x,y)$ in the middle plane. 
(f),(j) Thickness-averaged energy density profiles $\epsilon(x,y)$, obtained by integrating the energy density along the film thickness and normalizing by the thickness, which highlight regions with favorable (negative) and unfavorable (positive) twist contributions, including the outer shell. 
(g),(k) Preimages for two representative radii, $R=7$ and $R=22$, demonstrating the penetration of the conical phase into the hopfion interior in the form of additional loops.}
\end{figure*}

\begin{figure*}
  \centering
  \includegraphics[width=0.99\linewidth]{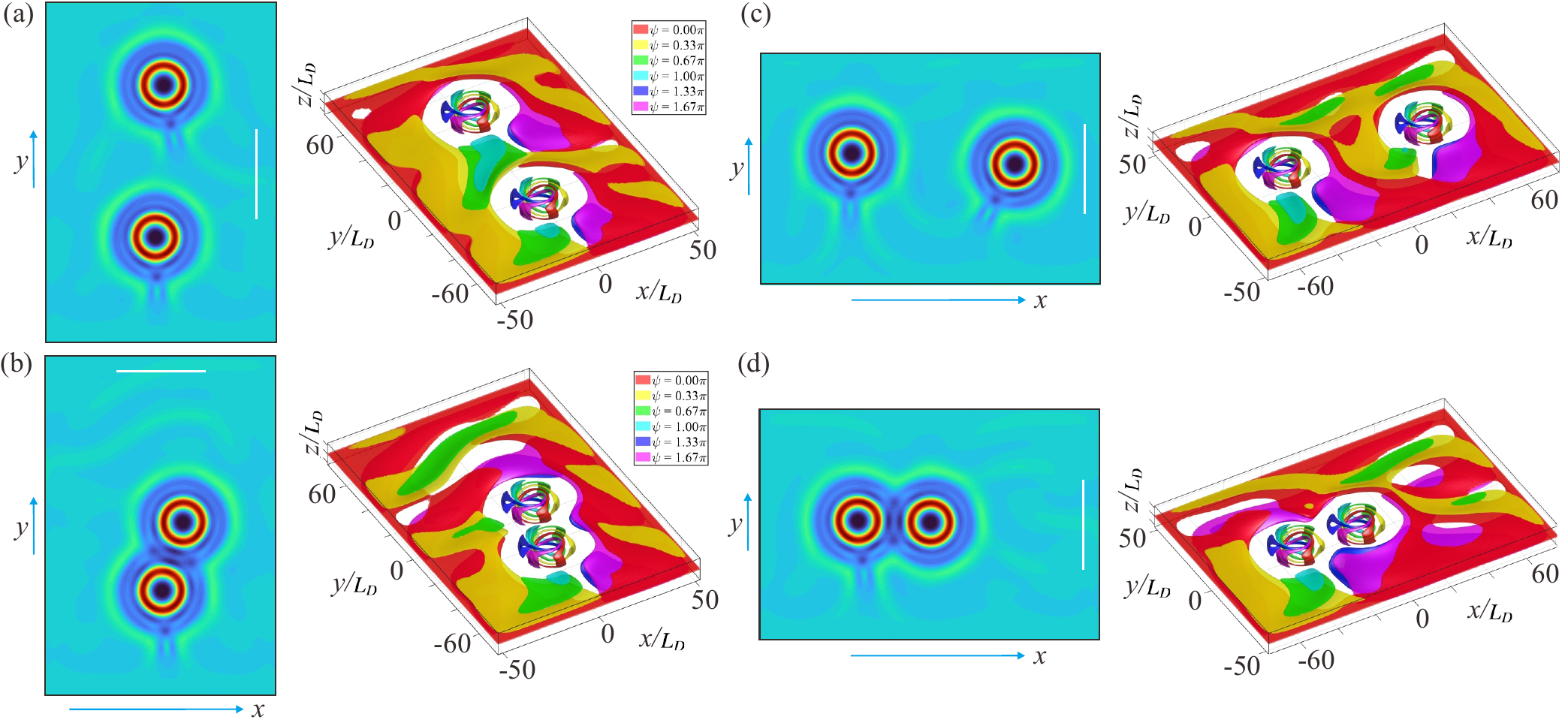}
\caption{\label{fig07} Representative configurations of two hopfions embedded in the conical state. (a),(c) Initial states with two remotely separated hopfions arranged along the $y$ axis (a) and the $x$ axis (c), respectively. After relaxation, the hopfions remain separated and tend to increase their mutual distance, indicating repulsive interaction at large separations. 
(b),(d) Relaxed bound states obtained for smaller initial inter-hopfion distances, demonstrating the formation of hopfion pairs for arrangements along $y$ (b) and $x$ (d). The pair configurations are essentially identical for both orientations, reflecting local rotation and strong frustration of the conical background, which is partially transformed into a CF-1–like state. 
Additional panels show the corresponding preimages, which do not reveal any extra linking compared to isolated hopfions. In selected color plots, a white line segment is included as a scale bar. The numerical values on the axes are omitted for clarity; the length of this segment corresponds to 40 (in the adopted length units), thereby providing the spatial scale of the image.}
\end{figure*}
%%%%%%%%%%%%%%%%%%%%%%%%%%%%%%%%%%%%%%%%%%%%%%%%%%%%%%%%%%%%%%%%%%%%%
%\subsection{Metastability of isolated hopfions within the conical phase; unstable hopfion lattices}

%%%%%%%%%%%%%%%%%%%%%%%%%%%%%%%%%%%%%%%%%%%%%%%%%%%%%%%%%%%%%%%%%%%%%%%
\section{Energetics and internal stability of hopfions within the conical phase \label{sect:Hopfstability}}

{\color{black}The two-dimensional cholesteric fingers of the second type considered above possess a positive eigen-energy with respect to the surrounding conical phase. The conclusions regarding isolated CF--2 fingers and their interactions therefore strictly apply to the quasi-one-dimensional geometry. % considered here, where the fingers are assumed to be translationally invariant along the $y$ direction. 
Extending a bimeron along the complementary in-plane direction further increases the total energy due to the additional frustration and modulation imposed on the surrounding conical phase. Consequently, in the thermodynamic limit of an infinitely extended quasi-two-dimensional system, periodic lattices of elongated CF--2 fingers are not expected to represent thermodynamically stable equilibrium phases within the parameter range considered in the present manuscript.

This conclusion does not exclude, however, the existence of finite-size metastable realizations under confinement. In experimental systems, the total excess energy of elongated fingers remains finite because the samples possess finite lateral dimensions and the fingers terminate at the sample boundaries. Such states may therefore survive as kinetically trapped or long-lived metastable configurations separated from the equilibrium conical phase by an energy barrier as depicted in Fig. \ref{fig03}(a).}

By contrast, upon closing onto itself and forming a hopfion, the bimeron acquires a finite total eigen-energy. The circularization of the finger eliminates the divergence associated with indefinite elongation and allows the system to partially optimize the balance between the negative-energy core and the positive-energy shells imposed by the surrounding conical state.

%However, by closing onto itself and forming a hopfion, the bimeron can render its eigen-energy finite and, moreover, reduce it by optimizing the circumference of the closed loop.

In the following sections, we analyze the properties of isolated hopfions with the Hopf index $|Q_H|=1$ and discuss the instability of the hopfion lattice. In doing so, we extensively rely on the terminology and the physical insights established for bimerons in the preceding sections.

%%%%%%%%%%%%%%%%%%%%%%%%%%%%%%%%%%%%%%%%%%%%%%%%%%%%%%%%%%%%%%%%%%%%%%%%%%%%
\subsection{Metastability of isolated hopfions within the conical phase}
%%%%%%%%%%%%%%%%%%%%%%%%%%%%%%%%%%%%%%%%%%%%%%%%%%%%%%%%%%%%%%%%%%%%%%%%%%%

Figure~\ref{fig06}(a) shows the eigen-energy $W(R)$ of a hopfion as a function of its radius~$R$ for several values of the magnetic field at $k_u^s = 20$ [see also the phase diagram in Fig. \ref{fig01}(b)]. The numerical procedure used to obtain this dependence is described in detail in Refs.~\cite{hopfions,leonov2026precursor} and is therefore only briefly summarized here. 
In short, the magnetization along a circular contour of radius~$R$ was pinned while evaluating the hopfion eigen-energy $W(R)$ with respect to the conical phase. In this manner, a set of hopfion configurations with different radii was generated and their energies were compared. As a consistency check, the pinning constraints were subsequently removed to verify that the relaxed configurations either retain their hopfion character and relax to the energy minimum, or collapse into torons in cases where the corresponding energy curve exhibits no minimum. 
As an additional constraint, the conical phase was fixed within a narrow region near the boundaries of the numerical grid with periodic boundary conditions in order to stabilize the background state. Specifically, the area outside the dashed black rectangle in Fig.~\ref{fig06}(b), corresponding to the conical phase, was “frozen”. This procedure is intended to suppress distortions of the conical spiral in the form of relative phase shifts, as discussed above for the CF--1 phase.

According to Fig.~\ref{fig06}(a), a well-defined energy minimum exists within a finite magnetic-field interval that extends beyond the critical field at which the bimeron lattice loses its local energy minimum [see Fig.~\ref{fig03}(a), insets].  %In particular, the hopfion minimum disappears slightly above $h=0.06$, which corresponds to the field at which the bimeron lattice itself becomes unstable. 
This observation once again illustrates the general principle that the metastability region of isolated hopfions—being circular derivatives of bimerons—is closely related to the stability range of the corresponding fingers. 
At $h = 0.09$, however, no energy minimum is present, indicating the instability of the isolated hopfions in this regime.
{\color{black} The metastability region of isolated hopfions in the  phase diagram is indicated by the curvilinear region $f-g-h$ in Fig. \ref{fig01}(b). }

%{\color{black} Here, we restrict ourselves to a representative cross-section of the phase diagram corresponding to the fixed value of the surface anisotropy parameter $k_u^s=20$. 
%
%A complete determination of the metastability regions of isolated hopfions in the full phase diagram would require extensive numerical simulations and variety of other methods  and is therefore beyond the scope of the present work. }

Figure~\ref{fig06} summarizes additional hopfion characteristics that help to clarify the origin of the energy minimum. First, we note that a circular hopfion must continuously adapt to the surrounding conical phase at each azimuthal angle, which results in a noticeably distorted structure. 
Color plots of the $m_z$ component of the magnetization at selected points along the energy curve for $h=0$ are shown in Figs.~\ref{fig06}(b), (c), (d), and (h) for the middle cross-section $z = 0$. These configurations exhibit a characteristic stick-like segment that enables the hopfion to match the rotating conical spiral. %The in-plane component of the magnetization within the cone is perpendicular to this segment and continuously rotates 
Figures~\ref{fig06}(b) and~\ref{fig06}(c) provide a three-dimensional view of the rotating magnetization, whereas Figs.~\ref{fig06}(d) and~\ref{fig06}(h) show the corresponding projections onto the $xy$ plane. 

The color plots in Figs.~\ref{fig06}(e) and~\ref{fig06}(i) display the energy density in the middle plane and allow one to distinguish different parts of the hopfion profile characterized by favorable and unfavorable twist of the magnetization. A clearer separation, however, is achieved by considering the averaged energy profiles $\epsilon(x,y)$ shown in Figs.~\ref{fig06}(f) and~\ref{fig06}(j). In this case, the energy density at each point with coordinates $(x,y)$ is integrated over the film thickness and subsequently normalized by the film width, yielding an effective two-dimensional energy distribution (not to be confused with the energy density $\varepsilon$ defined in Eq.~(\ref{epsilon})).

Within the eigen-energy minimum (red curve in Fig.~\ref{fig06}(a)), the averaged energy profile shown in Fig.~\ref{fig06}(f) can be decomposed into several distinct contributions. The hopfion core, characterized by a negative energy density, is surrounded by a ring of positive energy density, which is in turn followed by a second ring exhibiting negative energy density. Here, the terms ``negative'' and ``positive'' are used in a relative sense, i.e., with respect to the energy density of the surrounding conical phase, rather than in an absolute energetic sense. Finally, a narrow outer ring with small positive excess energy forms with respect to the surrounding conical state (referred to as the shell, in analogy with bimerons). 

The interplay between these contributions gives rise to the nonmonotonic dependence of $W(R)$ and determines the equilibrium radius.
When the hopfion radius decreases, the loss of negative energy density in the core is not compensated by the simultaneous reduction of the positive energy in the surrounding ring, resulting in an increase of the eigen-energy $W(R)$. Upon increasing the hopfion radius, $W(R)$ also grows, albeit for a different reason. In this regime, the conical phase partially penetrates into the hopfion core and reduces the magnitude of its negative energy density (Fig.~\ref{fig06}(j)). Moreover, although the area of the negative-energy ring expands, its favorable contribution is outweighed by the increasing positive energy of the intermediate ring and the outer shell, which also broaden with increasing~$R$.

Remarkably, the preimages plotted in Figs.~\ref{fig06}(g) and~\ref{fig06}(k) for two representative radii, $R=7$ and $R=21$, reveal that the conical phase manifests itself as additional loops penetrating into the hopfion interior. These loops reflect the partial invasion of the background conical modulation into the hopfion core as the radius increases. The preimages are constructed for $\theta = \pi/2$ and several values of the azimuthal angle $\psi$.

%%%%%%%%%%%%%%%%%%%%%%%%%%%%%%%%%%%%%%%%%%%%%%%%%%%%%%%%%%%%%%%%%%%%%%%%%%%%%%
\subsection{Hopfion pairing and cluster formation within the conical phase}

The characteristic energetic signature of isolated hopfions shown in Fig.~\ref{fig06}(f) suggests the presence of an attractive inter-hopfion potential. This conclusion follows from the same arguments as discussed above for bimerons: the positive energy contribution associated with the outer shells of isolated hopfions can be partially reduced when hopfions form clusters, owing to the effective overlap of these shells.

Moreover, the underlying mechanism responsible for the attractive inter-hopfion interaction within the conical phase closely parallels that established for isolated skyrmions. Conical-phase–mediated attraction between skyrmions has been theoretically analyzed in Refs.~\onlinecite{leonov2016three,loudon2018direct,capic2020skyrmion,leonov2022skyrmion}. Experimentally, clusters of such skyrmions were observed in thin (70\,nm) single-crystal samples of Cu$_2$OSeO$_3$ using transmission electron microscopy~\cite{loudon2018direct}. Similar behavior has also been reported in nematic liquid crystals, where skyrmions were shown to assemble into chain-like structures~\cite{ackerman2017}.

As an instructive example, we begin by analyzing the pairwise interaction between two hopfions.

%The internal structure of a hopfion pair is illustrated in Fig.~\ref{fig07}. 
%
In Fig.~\ref{fig07}, two hopfions are arranged along two perpendicular directions—either along $y$ (Figs.~\ref{fig07}(a),(b)) or along $x$ (Figs.~\ref{fig07}(c),(d)). In both cases, we initially place the hopfions far apart (Figs.~\ref{fig07}(a),(c)). After energy minimization using MuMax3, they remain well separated and even tend to increase their mutual distance, indicating a repulsive interaction at large separations. For smaller initial distances, however, the hopfions attract each other and form bound pairs (Figs.~\ref{fig07}(b),(d)). Remarkably, the pair configurations shown in Figs.~\ref{fig07}(b) and~\ref{fig07}(d) are essentially identical, indicating that the hopfions locally rotate the conical background until the global energy minimum is reached.

We therefore conclude that the interaction potential qualitatively exhibits the same shape as that obtained for bimerons in Fig.~\ref{fig05}(a): a local energy minimum corresponding to a bound state of coupled solitons, separated by an energy barrier from the configuration of two decoupled, isolated hopfions. A rigorous determination of the full two-dimensional interaction potential, however, would require extensive computations, since both hopfions must be pinned in the $xy$ plane at controlled relative positions. In addition, the conical phase must remain intact, which is technically challenging to ensure. 
This task is left for future work. In the present study, we therefore restrict ourselves to several representative cluster configurations.

\begin{figure}
  \centering
  \includegraphics[width=0.9\linewidth]{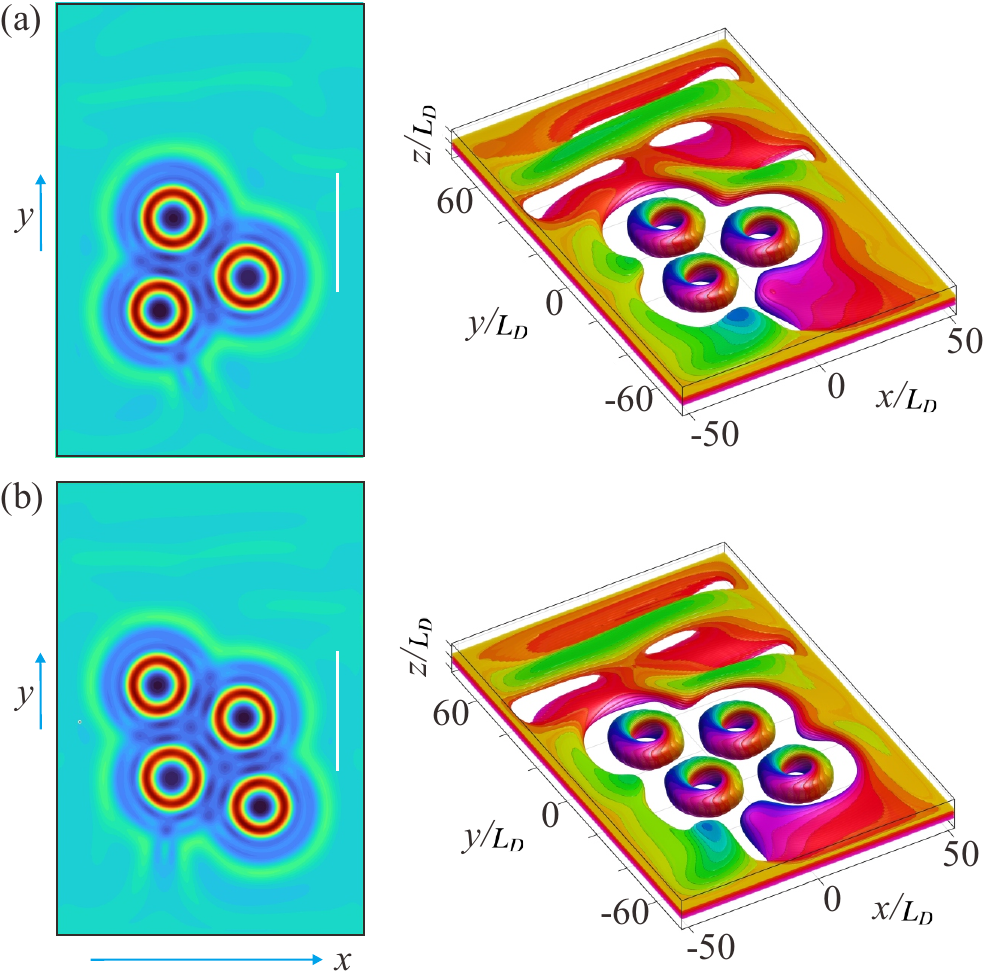}
\caption{\label{fig08}
Clusters of hopfions embedded in the conical phase after energy minimization. 
(a) Three hopfions relax into an equilateral triangular configuration, reflecting the tendency toward locally hexagonal ordering. 
(b) Four hopfions, initialized in a square arrangement, reorganize upon relaxation into a configuration composed of two adjacent triangles, demonstrating the preference for triangular coordination over square symmetry.}
\end{figure}

Hopfion clusters composed of more than two solitons exhibit a clear tendency toward hexagonal ordering. In particular, three hopfions (Fig.~\ref{fig08}(a)) relax into an equilateral triangular configuration, reflecting the effective isotropy of the attractive interaction at intermediate distances. 
For four hopfions (Fig.~\ref{fig08}(b)), even when starting from a square initial arrangement, the relaxed configuration does not preserve the square symmetry. Instead, the system reorganizes into two adjacent triangles, consistent with the preference for locally triangular coordination. This behavior indicates that the effective interaction favors sixfold packing, analogous to the hexagonal ordering observed in cluster phases of attracting skyrmions and other solitonic systems with short-range attraction and long-range repulsion.
Such a tendency suggests that larger hopfion aggregates may form compact clusters with locally triangular symmetry rather than square or stripe-like arrangements, provided that the conical background remains stable.

\begin{figure}
  \centering
  \includegraphics[width=0.99\linewidth]{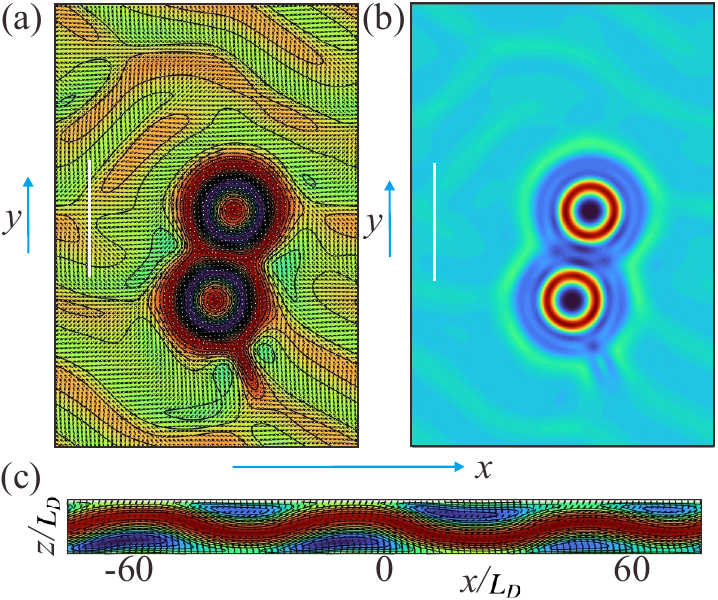}
\caption{\label{fig11}
Coupled hopfion pair in a frustrated conical background without boundary pinning. 
(a) Color plot of the $m_z$ component showing the maze-like domain pattern arising from spatial variations of the conical polar angle $\theta(x,y,z)$. 
(b) Thickness-averaged energy density $\epsilon(x,y)$, illustrating the outer positive-energy shells of the hopfions and the emergence of protuberances directly connected to these shells due to conical-domain formation. 
(c) $xz$ cross-section at the boundary of the numerical grid, displaying the undulating conical phase under periodic boundary conditions.
}
\end{figure}

The described mechanism of attractive interaction between hopfions remains operative even when the conical phase becomes strongly frustrated and does not preserve a uniform phase. Numerically, we remove the pinning of the conical state near the boundaries of the computational domain and retain only periodic boundary conditions. The resulting configuration for a coupled hopfion pair is shown in Fig.~\ref{fig11}. 
In this case, the polar angle of the cone becomes dependent on all Cartesian coordinates, $\theta(x,y,z)$, giving rise to a maze-like domain pattern in the color plot of the $m_z$ component (Fig.~\ref{fig11}(a)). Nevertheless, the color map of the thickness-averaged energy density $\epsilon(x,y)$ retains the same qualitative character as discussed above. The presence of conical domains manifests itself in the formation of “protuberances” of positive energy density directly connected to the outer shell (Fig.~\ref{fig11}(b)). This behavior suggests that isolated hopfions may interact already at larger separations and subsequently entangle into extended conglomerates, although this scenario requires further investigation. Figure~\ref{fig11}(c) shows the undulating conical phase in the $xz$ cross-section near the boundary of the numerical grid.

 %In addition, hopfions strongly frustrate the conical phase in their vicinity, locally transforming it into a CF-1–like state. 
%
%Through the formation of first-type fingers, isolated hopfions may begin to sense each other at larger separations and subsequently entangle into a bound pair. The preimages presented as additional panels in Fig.~\ref{fig07} do not exhibit any deviations from the standard preimages of isolated hopfions in Fig. \ref{fig06}; in particular, no additional linking has been detected. 

\begin{figure*}
  \centering
  \includegraphics[width=0.99\linewidth]{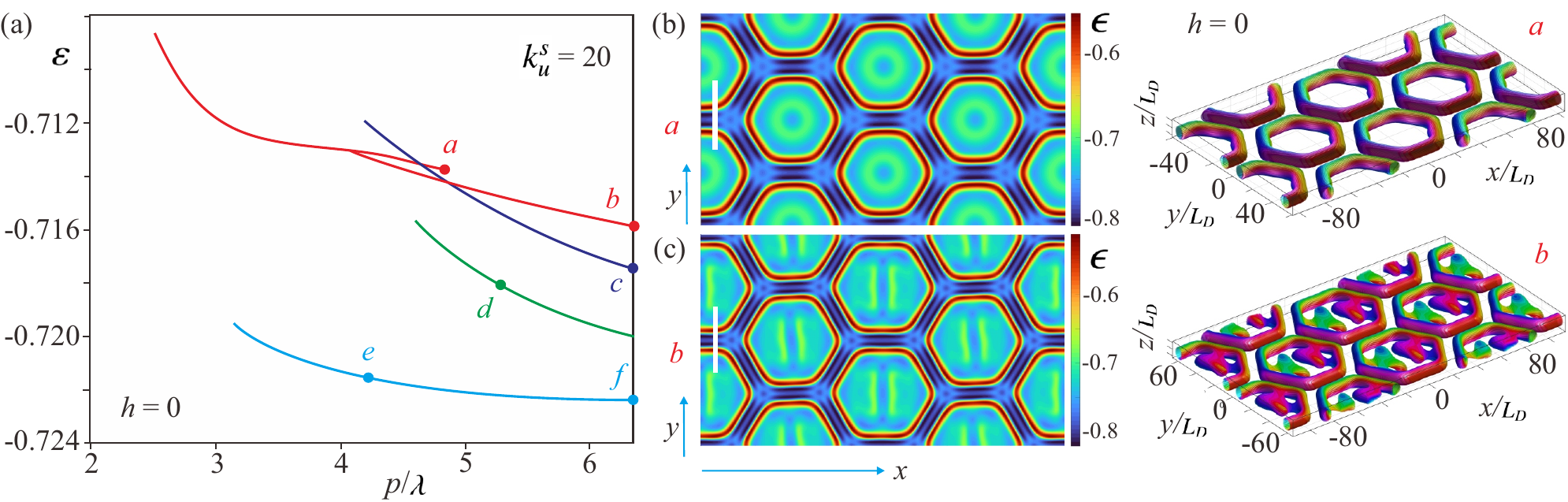}
\caption{\label{fig09}
(a) Energy density $\varepsilon$ of various hexagonal hopfion lattices as a function of the lattice period at $k_u^s=20$. 
The red curve corresponds to the hexagonal lattice of $Q_H=1$ hopfions and does not exhibit an energy minimum; branching reflects different modes of penetration of the conical phase into the hopfion interiors. 
The blue curve represents the “flower” state in which hopfions of approximately equilibrium size are surrounded by six CF--1 petals. 
The green curve corresponds to the hexagonal arrangement of hopfion bags, where a trivial hopfion ($Q_H=0$) encloses a nontrivial hopfion ($Q_H=1$). 
The light-blue curve shows the hexagonal lattice composed solely of trivial ($Q_H=0$) hopfions derived from the CF--1 phase, whose lattice period expands monotonically, indicating melting of the lattice. 
Points $a$–$f$ mark configurations shown in Figs.~\ref{fig09}(b),(c) and Fig.~\ref{fig10}.  
(b),(c) Spin configurations and corresponding preimages for two representative lattice periods (points $a$ and $b$), illustrating different patterns of conical-phase invasion into the hopfion cores during lattice expansion.
}
\end{figure*}

%%%%%%%%%%%%%%%%%%%%%%%%%%%%%%%%%%%%%%%%%%%%%%%%%%%%%%%%%%%%%%%%%%%%%%%%%%%%%
\subsection{Hopfion lattices}
%%%%%%%%%%%%%%%%%%%%%%%%%%%%%%%%%%%%%%%%%%%%%%%%%%%%%%%%%%%%%%%%%%%%%%%%%%%%%%

As a next step, in close analogy with the bimeron case, we investigate the stability of a hexagonal hopfion lattice. Hopfions are initially arranged in a perfectly hexagonal pattern, with the numerical grid chosen to preserve the corresponding lattice symmetry. The energy density $\varepsilon$ of the hexagonal lattice is then minimized with respect to the lattice period using \textsc{MuMax3}. The resulting energy dependence is summarized in Fig.~\ref{fig09}(a).

The red curve in Fig.~\ref{fig09}(a), corresponding to the hexagonal hopfion lattice, does not exhibit a local energy minimum. Instead, it shows branching behavior depending on how the conical phase penetrates the hopfion interiors. In the configuration corresponding to point~$a$, the conical phase appears as a circular domain at the center of each hopfion (Fig.~\ref{fig09}(b)). A more energetically favorable configuration is obtained at point~$b$, where the conical phase develops a more intricate structure inside the hopfions (Fig.~\ref{fig09}(c)). The corresponding preimages in Figs.~\ref{fig09}(b) and~\ref{fig09}(c) allow one to follow in detail the expansion of the lattice and the progressive invasion of the conical state.

In this respect, the energy-density curve closely resembles that of the bimeron lattice near the critical field at which the local energy minimum disappears, indicating the absence of a thermodynamically stable hopfion lattice despite the metastability of isolated hopfions or existence of hopfion clusters.

During the relaxation procedure, the initially imposed hopfion lattice may transform into energetically more favorable configurations. However, none of these states develops a local minimum of the energy density as a function of the lattice period (Fig. \ref{fig09}(a)).

\begin{figure*}
  \centering
  \includegraphics[width=0.9\linewidth]{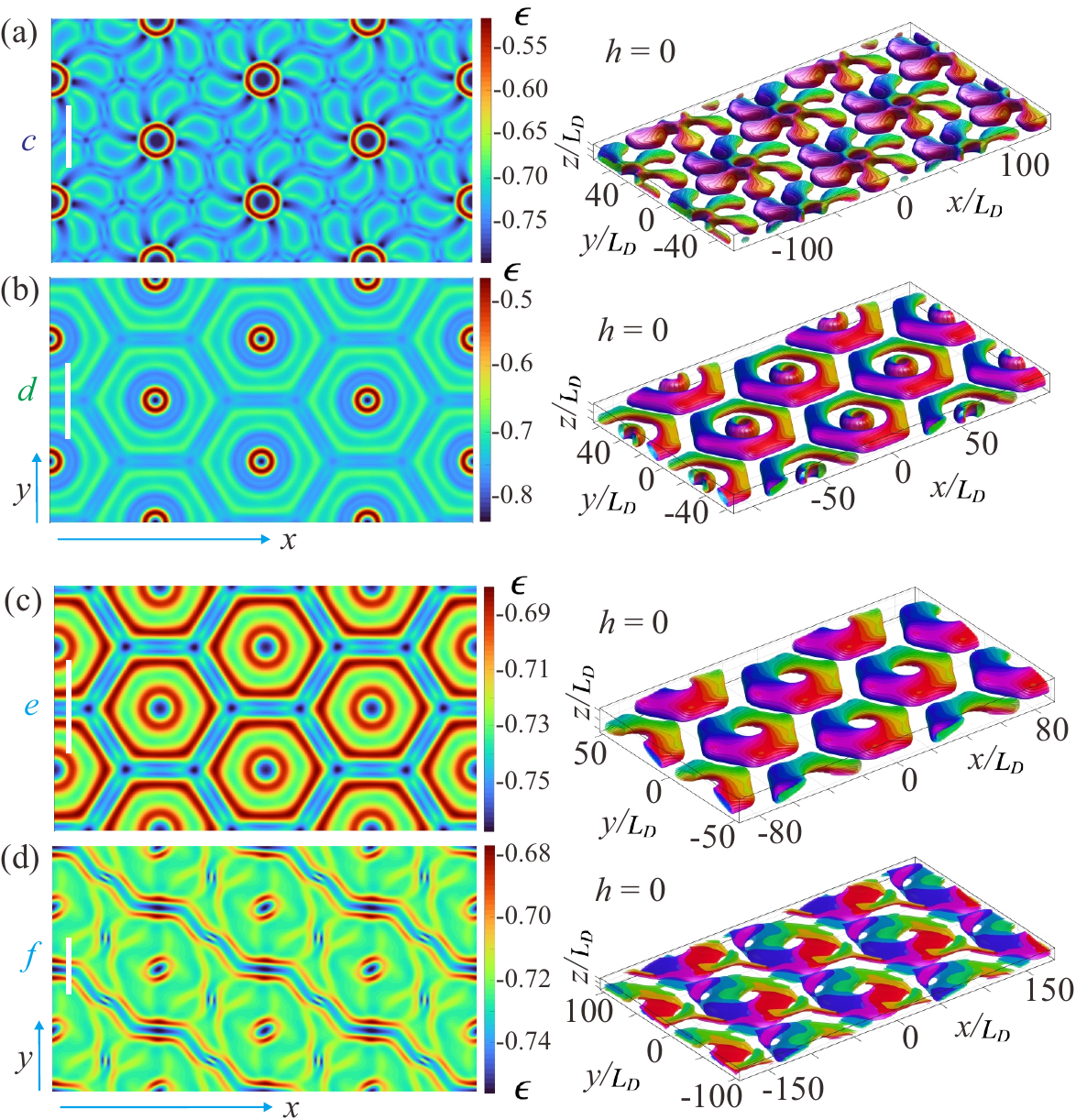}
\caption{\label{fig10}
Representative configurations of alternative hexagonal states derived from the hopfion lattice. 
(a) “Flower” state (point $c$ in Fig.~\ref{fig09}(a)), in which each $Q_H=1$ hopfion is surrounded by six CF--1 petals formed by first-type fingers filling the inter-hopfion space. 
(b) Hexagonal arrangement of hopfion bags (point $d$ in Fig.~\ref{fig09}(a)), where a trivial hopfion ($Q_H=0$) encloses a nontrivial hopfion ($Q_H=1$); color plots of the thickness-averaged energy density $\epsilon(x,y)$ and the corresponding preimages illustrate the internal structure of this composite phase. 
(c),(d) Configurations corresponding to points $e$ and $f$ in Fig.~\ref{fig09}(a), showing the progressive expansion and melting of the hexagonal lattice composed of trivial ($Q_H=0$) hopfions derived from the CF--1 phase, driven by the tendency of the CF--1 background to occupy space continuously rather than forming closed circular objects.
}
\end{figure*}

Given that the conical phase can be readily replaced by the CF--1 phase, isolated hopfions may retain approximately the same equilibrium radius as in finite clusters. In this case, instead of continuously increasing their radii—as observed along the red curve in Fig.~\ref{fig09}(a)—the additional space between neighboring hopfions becomes filled with fingers of the first type, forming six CF--1 “petals” around each hopfion (point~$c$ in Fig.~\ref{fig09}(a) and the corresponding spin configuration in Fig.~\ref{fig10}(a)). 
This “flower” state, however, is likewise unstable and does not exhibit any energy minimum (blue curve in Fig.~\ref{fig09}(a)). The reason is the energetic advantage of the CF--1 phase, which tends to expand and occupy the entire inter-hopfion region, thereby preventing the formation of a stable periodic hopfion lattice.

The CF--1 phase may also intervene in the form of circular hopfions with Hopf index $Q_H = 0$, i.e., as closed, swirled first-type fingers. The driving mechanism for the emergence of this phase is the same as in the preceding case, namely, the tendency of the system to fill the interstitial space with the energetically favorable CF--1 phase. 
Figure~\ref{fig10}(b) illustrates the internal structure of this state by means of color plots of the thickness-averaged energy density $\epsilon(x,y)$ and the corresponding preimages for the configuration labeled by point~$d$ in Fig.~\ref{fig09}(a). This phase can be interpreted as a hexagonal arrangement of hopfion bags, in which a trivial hopfion ($Q_H = 0$) encloses a nontrivial hopfion with $Q_H = 1$. 
However, this phase likewise does not exhibit any energy-density minimum, as indicated by the green curve in Fig.~\ref{fig09}(a).

We have also analyzed a hexagonal hopfion lattice composed of trivial hopfions derived from the CF--1 phase. This configuration, however, does not represent an equilibrium state with a well-defined lattice period. Instead, the lattice period gradually increases, as illustrated by the light-blue curve in Fig.~\ref{fig09}(a).
The configurations corresponding to points~$e$ and~$f$ in Fig.~\ref{fig09}(a) are shown in Figs.~\ref{fig10}(c) and~\ref{fig10}(d), respectively. These panels allow one to closely follow the melting process of this hopfion lattice, which is driven by the energetic preference of the CF--1 phase to occupy the available space continuously rather than forming closed, circular hopfion structures.

All examples of hexagonally ordered hopfion states discussed above illustrate the following general point: although a given configuration may exist as a stationary solution within a numerical grid of prescribed dimensions, this does not imply the existence of a thermodynamically stable phase with a well-defined equilibrium lattice period. In other words, numerical realizability under constrained boundary conditions is not equivalent to the presence of a true energy minimum with respect to lattice spacing.

This consideration may be relevant to the theoretically reported stable hopfion lattice in Ref.~\cite{tai2018}, where no explicit minimization of the energy density with respect to the lattice period was presented. Moreover, the images of the skyrmion lattices shown in Fig.~S2 of the Supplementary Information of Ref.~\cite{tai2018} display the same number of skyrmions within an identical area for different values of the magnetic field. This observation suggests that the lattice period may not have been systematically optimized with respect to the applied field.
In this sense, the methodology employed in the present work is instrumental in distinguishing a genuinely stable hopfion lattice from configurations that arise as numerical artifacts of constrained boundary conditions.

The present methodology is likewise instrumental for the analysis of experimental reports on stable hopfion lattices. In particular, such an analysis should begin with the character of the pairwise interaction between hopfions. The results of the present work, as well as those of Ref.~\cite{hopfions}, predict an attractive interaction mediated by the conical phase, whereas a repulsive interaction is expected within the homogeneous state.

In the former case, the attractive interaction enables the formation of extended hopfion clusters. Experimentally, hopfions may be created and manipulated, for instance, by focused laser tweezers in CLC, allowing controlled adjustment of the distance between two solitons. The formation of clusters, however, does not imply the existence of a hopfion crystal. % or a thermodynamically stable lattice.

In the latter case, the repulsive interaction prevents cluster formation. If several hopfions are nucleated experimentally, they will progressively move apart from each other. Moreover, if hopfions acquire a lower energy relative to the homogeneous state, they would tend to elongate into an extended finger phase and eventually fill the available space. Even if an ideal hopfion lattice could be created, it remains unclear which mechanism would suppress the elongation instability. Indeed, a modulated finger phase can tile space with unit efficiency, whereas a hopfion lattice necessarily contains extended regions of voids.  

%“vacuum,” both inside the hopfion cores and in the interstitial areas between neighboring objects. This geometric inefficiency promotes deformation of the hopfion textures, leading to the collapse of their vacuum cores and subsequent elongation into finger-like structures.

Recently, a Wigner crystal of hopfions in chiral liquid-crystal colloidal magnets, reported to be stable at magnetic fields of opposite orientations, was described in Ref.~\cite{ackerman2017static}. However, these experimental findings warrant closer examination in light of the questions and principles outlined above.

%%%%%%%%%%%%%%%%%%%%%%%%%%%%%%%%%%%%%%%%%%%%%%%%%%%%%%%%%%%%%%%%%%%%%%%%%%%%
\section{Conclusions}

The results of this work establish a coherent physical picture of soliton behavior in thin films of chiral magnets and chiral liquid crystals hosting a conical background. Both bimerons (CF--2 fingers) and hopfions possess positive eigen-energy relative to the conical phase. Nevertheless, they experience an effective attractive interaction mediated by distortions of the surrounding conical state. The origin of this attraction can be traced to the shell structure accompanying isolated solitons: the outer regions of positive excess energy partially overlap when two solitons approach each other, thereby reducing the total energy of the pair. This mechanism closely parallels the conical-phase–mediated attraction previously established for isolated skyrmions.

For bimerons, the attractive inter-soliton potential exhibits a local minimum separated by an energy barrier from the configuration of two isolated objects. Although isolated bimerons attract and form bound states or chains, the corresponding bimeron lattice loses its equilibrium lattice period once the critical field is exceeded. Thus, cluster formation persists beyond the stability range of the periodic phase, demonstrating that pairwise attraction alone does not guarantee lattice stability.

Upon circularization into hopfions, the eigen-energy becomes finite and displays a nonmonotonic dependence on the hopfion radius. A metastability window emerges within a finite magnetic-field interval closely linked to the stability region of the underlying bimerons. The equilibrium hopfion radius results from a delicate balance between negative-energy core regions and surrounding positive-energy rings and shells imposed by the conical background. At larger radii, penetration of the conical phase into the hopfion interior progressively diminishes the energetic advantage of the core, ultimately eliminating the energy minimum.

Importantly, isolated hopfions also exhibit attractive interaction within the conical phase and form bound pairs and compact clusters with a pronounced tendency toward local hexagonal ordering. However, despite this attractive pair potential, a thermodynamically stable hexagonal hopfion lattice does not emerge. Explicit energy minimization with respect to the lattice period reveals the absence of a global minimum. Instead, the system evolves toward configurations in which either the conical phase or the energetically favorable CF--1 phase progressively occupies the inter-hopfion space. Alternative states, including flower-like configurations and hexagonal arrangements of hopfion bags, likewise fail to develop an equilibrium lattice period.

These findings establish a regime of attraction without crystallization: solitons attract and cluster, yet long-range periodic order is suppressed by energetic competition between localized topological objects and extended modulated phases. More generally, our results highlight the decisive role of background frustration and phase competition in governing the stability of three-dimensional topological solitons in confined geometries. The interplay between topology, confinement, and conical-phase restructuring provides a versatile framework for tuning soliton interactions and offers guiding principles for the controlled creation of metastable soliton clusters in chiral-magnet thin films and related systems.

\section{Acknowledgments}
A.O.L.  acknowledges support by JSPS KAKENHI Grant Number 26K06945 (Grant-in-Aid for Scientific Research (C)).

\bibliographystyle{apsrev4-2}
\bibliography{refs}

\end{document}